\documentclass[fleqn,usenatbib]{mnras/mnras}

\usepackage{newtxtext,newtxmath,cases,systeme,soul}

\usepackage[T1]{fontenc}

\DeclareRobustCommand{\VAN}[3]{#2}
\let\VANthebibliography\thebibliography
\def\thebibliography{\DeclareRobustCommand{\VAN}[3]{##3}\VANthebibliography}

\usepackage{graphicx}	
\usepackage{amsmath}	
\usepackage{hyperref}

\title[A new tool for power-law distributions]{A new tool to derive simultaneously exponent and extremes of power-law distributions}

\author[S. Pezzuto et al.]{S. Pezzuto,$^{1}$\thanks{E-mail: stefano.pezzuto@inaf.it}
A. Coletta,$^{1,2}$
R. S. Klessen,$^{3,4}$
E. Schisano,$^{1}$
M. Benedettini,$^{1}$
D. Elia,$^{1}$
S. Molinari,$^{1}$\newauthor
J. D. Soler,$^{1}$
A. Traficante,$^{1}$
\\
$^{1}$INAF - IAPS, via Fosso del Cavaliere, 100, I-00133 Roma, Italy\\
$^{2}$Dipartimento di Fisica, Università di Roma La Sapienza, Piazzale Aldo Moro 2, I-00185, Roma, Italy\\
$^{3}$Universität Heidelberg, Zentrum für Astronomie, Institut für Theoretische Astrophysik, Albert-Ueberle-Str 2, D-69120 Heidelberg, Germany\\
$^{4}$Universität Heidelberg, Interdisziplinäres Zentrum für Wissenschaftliches Rechnen, Im Neuenheimer Feld 205, D-69120 Heidelberg, Germany}

\date{Accepted XXX. Received YYY; in original form ZZZ}

\pubyear{2015}

\begin{document}
\label{firstpage}
\pagerange{\pageref{firstpage}--\pageref{lastpage}}
\maketitle

\begin{abstract}
Many experimental quantities show a power-law distribution $p(x)\propto x^{-\alpha}$. In astrophysics, examples are: size distribution of dust grains or luminosity function of galaxies. Such distributions are characterized by the exponent $\alpha$ and by the extremes $x_\text{min}$ $x_\text{max}$ where the distribution extends. There are no mathematical tools that derive the three unknowns at the same time. In general, one estimates a set of $\alpha$ corresponding to different guesses of $x_\text{min}$ $x_\text{max}$.  Then, the best set of values describing the observed data is selected a posteriori. In this paper, we present a tool that finds contextually the three parameters based on simple assumptions on how the observed values $x_i$ populate the unknown range between $x_\text{min}$ and $x_\text{max}$ for a given $\alpha$. Our tool, freely downloadable, finds the best values through a non-linear least-squares fit. We compare our technique with the maximum likelihood estimators for power-law distributions, both truncated and not. Through simulated data, we show for each method the reliability of the computed parameters as a function of the number $N$ of data in the sample. We then apply our method to observed data to derive: i) the slope of the core mass function in the Perseus star-forming region, finding two power-law distributions: $\alpha=2.576$ between $1.06\,M_{\sun}$ and $3.35\,M_{\sun}$, $\alpha=3.39$ between $3.48\,M_{\sun}$ and $33.4\,M_{\sun}$; ii) the slope of the $\gamma$-ray spectrum of the blazar J0011.4+0057, extracted from the Fermi-LAT archive. For the latter case, we derive $\alpha=2.89$ between 1,484~MeV and 28.7~GeV; then we derive the time-resolved slopes using subsets 200 photons each.
\end{abstract}

\begin{keywords}
Methods: data analysis – Methods: statistical – Stars: luminosity function, mass function – Submillimeter: general – Quasars: individual: J0011.4+0057 – Gamma rays: galaxies
\end{keywords}



\section{Introduction}
An observable is said to follow a power-law distribution if the probability to measure a value between $x$ and $x+\text{d}x$ is
\begin{equation}
p(x)\text{d}x=Cx^{-\alpha}\text{d}x\label{px}
\end{equation}
where $C$ is a normalization factor and $\alpha>1$ is the exponent. 
Alternatively, we can define the distribution as
\begin{equation}
n(x)\text{d}x=Np(x)\text{d}x\propto x^{-\alpha}\text{d}x \label{nx}
\end{equation}
where $n(x)$ is the number of events with value between $x$ and $x+\text{d}x$ and $N$ is the total number of events. The study of this distribution dates back to the end of 19th century when the Italian mathematician Vilfredo Pareto (1848 - 1923) first introduced the power-law distribution in the context of economy studies \citep{clark}. The distribution can be defined over the interval $x_\text{min}\le x<\infty$, or, for a \textit{truncated} power law, over the closed interval $x_\text{min}\le x<x_\text{max}$\footnote{The reason why $x<x_\text{max}$ and not $x\le x_\text{max}$ is because the probability to observe a specific value $x$ is zero for a continuous distribution. That’s why $p(x)$ is always defined over an interval between $x$ and $x+\text{d}x$. As a consequence, $p(x_\text{max})$ can be different from zero only over an interval between $x_\text{max}$ and $x_\text{max}+\text{d}x$, that would imply the possibility to observe values $x>x_\text{max}$. This is also the reason why in a uniform distribution a random variable $r$ is defined over the interval $0\le r<1$. Note also that when $\alpha<0$ the distribution must be bounded by $x_\text{max}$: $0<x\le x_\text{max}$.}.

In astronomy we have many examples of such distributions. The number $n$ of the interstellar dust grains with size $a$ follows the law $n(a)\propto \alpha^{-3.5}$ \citep{MRN}; the stellar initial mass function (IMF), the number of stars in the neighborhood of the Sun having mass between $m$ and $m+\text{d}m$, is described in terms of a log-normal function followed by a power law \citep{chabrier} with $\alpha=2.35$, an exponent first found by \citet{salpeter}; or with two or three power laws \citep{kroupa_jerabkova_2021,kroupa2001}\footnote{Paper by Kroupa \& Jerabkova~(2021) available at \url{https://arxiv.org/pdf/2112.10788.pdf}.}, one of which has Salpeter’s exponent. The precursor of the IMF is the pre-stellar core mass function (CMF): a pre-stellar core is a condensation of gas and dust that is collapsing or expected to collapse because its self-gravity is not supported by pressure (thermal or magnetic). The CMF in star-forming regions is similar in shape to the IMF \citep[e.g.,][and references therein]{perseo}, so one or more power laws are expected \citep[not necessarily with the same slope of the IMF, e.g.,][]{motte}.

The probability density function (PDF) of H$_2$ column density maps is an important diagnostic tool to disentangle the different physical mechanisms acting in the star-formation clouds. At high
densities, the observed PDF is modelled with a power-law tail \citep[e.g.,][]{nicola}. Deriving the slope $\alpha$ of this tail is very important because, from simulations, \citet{fedKle} derived a relation between $\alpha$ and the star formation efficiency, i.e., the percentage of mass actively forming stars, of a cloud.

Moving to cosmology, other examples are the luminosity function of the galaxies, that follows a power law in the low-luminosity regime \citep{schechter}, or the spectrum of the primordial density perturbation \citep{planck}\footnote{In this case, however, the function distribution is more complex because $\alpha$ is not constant, rather $\alpha=\alpha(x)$.}.

The electromagnetic spectrum of a source is also a distribution function of photons and, depending on the physical mechanism responsible for the emission or absorption, the continuum can be expressed as a power law \citep{RL}. The most obvious example of power-law emission is the blackbody radiation at temperature $T$ in the Rayleigh–Jeans regime: $B_\nu(T)\text{d}\nu\propto\nu^2\text{d}\nu$.

There is a number of approaches to derive $\alpha$ and a summary can be found in \citet{olmi}: roughly speaking, they can be divided into two classes. In one approach the number of values falling in a certain range of $x$ are counted and a histogram is built; then, Eq.~(\ref{nx}) is fit to the distribution. There are many drawbacks with this approach. The limits of the distribution $x_\text{min}$ and $x_\text{max}$, unless known a priori, must be chosen with some arbitrariness; the single measurement $x_i$ cannot be used alone but we have to group all the measurements in bins. This implies that a bin size must be chosen, but potentially the shape of the histogram, and thus $\alpha$, can depend on the bin sizes. Only for high $N$ the bin size is not an issue anymore \citep{nicola2}. Furthermore, random fluctuations in the high-values tail of the histogram can generate zero-count bins which are incompatible with Eq.~(\ref{nx}).

To overcome some of these difficulties, \citet{maizUbe} proposed to use histograms with adaptive bin sizes, so that each bin contains the same number of events. In this way one avoids both zero-count bins and possible changes of shape of the histogram. However, this does not solve the problem of knowing over which interval can $\alpha$ be computed. In other words, the limits of the distribution remain arbitrary.

The second class of $\alpha$ estimators is based on using the single values $x_i$ instead of building a histogram; this is the case, for instance, when using the maximum likelihood estimator (MLE). Also with this approach, however, we do not have a way to derive directly from data $x_\text{min}$, or $x_\text{min}$ and $x_\text{max}$ for a truncated power law. One has to compute different values of $\alpha$ by varying $x_\text{min}$ and through a posteriori comparisons, for instance using a Kolmogorov-Smirnov statistics, one can find the optimal choice for the parameters \citep{velt}.

\citet{aban} derived the equation that gives the MLE of $\alpha$ in the case of a truncated power law using the minimum and the maximum value of the observed data to estimate the parameters $x_\text{min}$ and $x_\text{max}$. They demonstrated that this choice is asymptotically valid, in the sense that the observed minimum and maximum are indeed equal to $x_\text{min}$ and $x_\text{max}$ when the number of observed data $N\rightarrow\infty$. The problem with this approach is that generally data do not follow a power-law distribution over the entire range of observed values, so that the observed minimum and maximum can be very different from $x_\text{min}$ and $x_\text{max}$. To deal with this problem, an exponential truncated power law, a distribution of the form $p(x)\propto x^{-\alpha}\times\exp(-ax)$ is also used, with $a$ some scale factor. The exponential part forces the power law to go to zero, at low or high values \citep[see][for $\gamma$-ray spectra]{FLGC}.

\citet{MK2009} presented alternative methods to MLE, included a modified MLE, and compared the performance of each technique. They also provided estimators of $x_\text{max}$ under the hypothesis that $x_\text{max}>\text{max}(x)$, i.e., $x_\text{max}$ cannot be smaller than the maximum of the observed data.

In this paper we propose a new tool that solves the problem of computing $\alpha$, $x_\text{min}$ and $x_\text{max}$ at the same stage, directly from the data. We start making a model of how the data will distribute most frequently given a set of parameters $\{\alpha,x_\text{min},x_\text{max}\}$, under the hypothesis that the intrinsic distribution is a power law; this allows us to define $N$ intervals, where $N$ is the number of observed values, each containing, on average, one value. Then, for each interval, we derive the expectation values that are compared with the observed values. The best-fit solution of our problem is found by varying the parameters through the well-known least-squares technique. Having to deal with uncertainties in the observable, our tool allows solutions with $x_\text{max}<\text{max}(x)$, differently from \citet{MK2009}; also, we leave open the possibility that the power law does not extend up to $\text{max}(x)$, which implies necessarily that $x_\text{max}<\text{max}(x)$.

The paper is organized as follows: in Section~\ref{mD} we derive the constant $C$ in equation~(\ref{px}), then we present our method to compute $\alpha$, $x_\text{min}$ and $x_\text{max}$ (Section~\ref{sec:LST}); afterwards, in Section~\ref{HillE} the MLE solution is introduced and compared with the Hill's estimator, while in Section~\ref{sceltaX} we show the consequence of adopting a wrong $x_\text{min}$ and the consequence of assuming that $x_\text{max}$ is not finite when it actually is. Most of the material in Section~\ref{HillE} is already known and available in literature. It is here reported for reader's convenience.

In Section~\ref{secSim} we present a set of simulations aimed at deriving the asymptotic behaviour of all the three methods, and their reliability when small samples of data are available. We consider the case of unlimited $x_\text{max}$ (Section~\ref{simOSB}) and finite $x_\text{max}$ (Section~\ref{simOTB}). After the simulations we apply our method to observed data presenting the general strategy in Section~\ref{realD}; then, we use our technique to derive the CMF in Perseus based on \textsl{Herschel} data \citep{perseo} in Section~\ref{CMFP}, showing that even with few cores it is possible to describe the data with two power-law distributions, while in Section~\ref{Blazar} our method is used to measure the slope of the $\gamma$-ray spectrum of the flat-spectrum radio quasar J0011.4+0057, with data extracted from the \textsl{Fermi}-LAT fourth source catalog \citep{FLGC}. Conclusions are reported in Section~\ref{conclusioni}.

Some additional mathematical material is given in Appendix~\ref{addForm} and Appendix~\ref{ancForm}: in particular we derive the formulae for special cases of $\alpha$ for our method in Section~\ref{specCas}, and of the derivatives necessary to solve the non-linear least-squares problem in Section~\ref{derAlpha}. In Appendix~\ref{ancForm} we derive asymptotic values of some limits appearing in Section~\ref{sceltaX}. Finally, in Appendix~\ref{genera} we give the formulae to construct a power-law distribution starting from a uniform random numbers distribution.

\section{Mathematical derivations}\label{mD}
We start this section by deriving the constant $C$ in equation~(\ref{px}), exploiting the relation $\int p(x)\text{d}x=1$. For a truncated distribution, $x_\text{min}\le x<x_\text{max}$,
\begin{equation}
1=C\int_{x_\text{min}}^{x_\text{max}}x^{-\alpha}\text{d}x\rightarrow C=\frac{1-\alpha}{x_m^{1-\alpha}(f^{1-\alpha}-1)}\label{defC}
\end{equation}
where $x_M\equiv x_\text{max}$, $x_m\equiv x_\text{min}$ and $f=x_M/x_m$; equation~(\ref{defC}) is valid for all $\alpha$ but 1 when it becomes
\begin{equation}
C=\left[\ln\frac{x_M}{x_m}\right]^{-1}.
\end{equation}
Then
\begin{numcases}{p(x)\text{d}x=}
\frac{1-\alpha}{x_m(f^{1-\alpha}-1)}\left(\frac{x}{x_m}\right)^{-\alpha}\text{d}x\,\,\,\alpha\in\mathbb{R}-\{1\}, \label{pxt}
\\
\left[\ln\frac{x_M}{x_m}\right]^{-1}\frac{\text{d}x}x\,\,\,\,\,\,\alpha=1. \label{pxt1}
\end{numcases}
When $\alpha=0$, equation~(\ref{pxt}) becomes the uniform distribution.

If the distribution is not truncated ($x_\text{min}\le x<+\infty$, $\alpha>1$)
\begin{equation}
C=\frac{\alpha-1}{x_m^{1-\alpha}}\label{CM1}
\end{equation}
and
\begin{equation}
p(x)\text{d}x=\frac{\alpha-1}{x_m^{1-\alpha}}x^{-\alpha}\text{d}x.\label{pxnt}
\end{equation}


\subsection{The least-square approach}\label{sec:LST}
Suppose we have $N$ data $x_i$ that we know, or we think, follow a power-law distribution with exponent $\alpha$. We assume that, most frequently, the observed values will be distributed uniformly in intervals of equal probability; so, we look for $N-1$ values $\xi_i$ that divide the interval between $x_m$ and $x_M$ in $N$ sections such that
\begin{equation}
\int_{x_m}^{\xi_1}p(x)\text{d}x=\int_{\xi_1}^{\xi_2}p(x)\text{d}x=\cdots=\int_{\xi_{N-2}}^{\xi_{N-1}}p(x)\text{d}x=\frac1N\label{equPro}
\end{equation}
where $p(x)$ is given by equation~(\ref{pxt}). After solving the integrals one arrives at
\begin{equation}
\xi_i=x_m\left[1-\frac{i}N\left(1-f^{1-\alpha}\right)\right]^\frac1{1-\alpha}\,\,\,1\le i\le N-1,\,\,\alpha\ne1.\label{defXi}
\end{equation}

Now, for each interval we compute the expectation value $\int xp(x)\text{d}x/\int p(x)\text{d}x=N\int xp(x)\text{d}x$. After some algebra we derive
\begin{align}
\bar\xi_i=&\frac{\alpha-1}{2-\alpha}\frac{Nx_m}{1-f^{1-\alpha}}\left\{\left[1-\frac{i}N\left(1-f^{1-\alpha}\right)\right]^\frac{2-\alpha}{1-\alpha}+\right. \label{expV}\\
&-\left.\left[1-\frac{i-1}N\left(1-f^{1-\alpha}\right)\right]^\frac{2-\alpha}{1-\alpha}\right\}\,\,\,1\le i\le N,\,\,\alpha\ne1,\,\,\alpha\ne2\nonumber
\end{align}
or
\begin{equation}
\bar\xi_i=\frac{\alpha-1}{2-\alpha}\frac{Nx_m}{1-f^{1-\alpha}}\left[1-\frac{i}N\left(1-f^{1-\alpha}\right)\right]^\frac{2-\alpha}{1-\alpha}-\sum_{i=1}^{i-1}\bar\xi_i\label{altExpV}
\end{equation}
which is computationally more effective. The equations valid for $\alpha=1$ and $\alpha=2$ are given in Appendix~\ref{specCas}.

So, if we observe $N$ values $x_i$, first we sort them in increasing order obtaining the sequence $x_1<x_2<\cdots<x_N$; then, with an initial set of parameters $p_j=\{\alpha,x_m,x_M\}_{\text{ini}}$ (see below) we compute the expected values $\bar\xi_i$ using equation~(\ref{expV}). The $i$-th residual is
\begin{equation}
r_i=\frac{(x_i-\bar\xi_i)^2}{\sigma_i^2}
\end{equation}
where $\sigma_i$ is the uncertainty associated to $x_i$.

The least-squares best-fit set of $p_j$ minimizes, by definition, the sum of the residuals
\begin{equation}
\sum_{i=1}^N r_i=\sum_{i=1}^N\frac{(x_i-\bar\xi_i)^2}{\sigma_i^2}.
\end{equation}
The minimum is found by imposing the condition
\begin{equation}
\frac{\partial}{\partial p_j}\sum r_i=\frac{\partial\bar\xi_i}{\partial p_j}\frac{\partial}{\partial \bar\xi_i}\sum_{i=1}^N\frac{(x_i-\bar\xi_i)^2}{\sigma_i^2}=0
\end{equation}
or
\begin{equation}
-2\sum\frac{x_i-\bar\xi_i}{\sigma_i^2}\frac{\partial\xi_i}{\partial p_j}=0\Rightarrow\sum\frac{x_i-\bar\xi_i}{\sigma_i^2}\frac{\partial\xi_i}{\partial p_j}=0\label{lS}
\end{equation}
which is a non-linear system solved for the three unknown $p_j$. We report in Appendix~\ref{derAlpha} the derivatives $\partial\xi_i/\partial p_j$ in case they are required by the chosen algorithm to solve the system.

To solve the non-linear least-squares problem we use the routine \textsl{curve\_fit} from SciPy \citep{sciPy}. As initial estimates of the parameters we use the maximum likelihood estimator (MLE, see below) to derive an initial value for $\alpha$, while $x_1$, the smallest value of our set, and $x_N$, the largest value, are used to estimate $x_m$ and $x_M$. If the search of the best-fit solution is not constrained, the parameters can assume any value. As we deal with observed data, we are not necessarily interested in the best solution in a mathematical sense. For instance, we trust a solution for which $x_M$ is not much different from $x_N$; and, similarly, $x_m$ should be not too different from $x_1$. Moreover, depending on the specific problem, also $\alpha$ can be constrained within a restricted interval.

The found solution is the best description of our data set in terms of a power law and under the hypothesis of uniformity, as stated by the set of equations~(\ref{equPro}). Let us have a closer look at these equations.

We have divided the interval between $x_m$ and $x_M$ in $N-1$ intervals. All the equations $\int_{\xi_i}^{\xi_{i+1}}p(x)\text{d}x=\frac1N$ mean that in the interval between $\xi_i$ and $\xi_{i+1}$ we expect to observe one datum only. One may argue that this is nothing but a particular case of the variable bin size histograms technique \citep{maizUbe}, sizing each bin to contain one observed value. Our method, however, is very different from that because the variable bin size technique does not give a way to derive $x_m$ and $x_M$: they must be assumed a priori. Also, one should check a posteriori that changing the number of data per bin does not impact significantly the derived exponent. But the most important difference is that we build a model that is compared with the observed data, wheras in \citet{maizUbe}’s approach, the intervals are built around each value. In our approach, if the set of $\{x_i\}$ does not follow a power law, this will be witnessed either by high values of $\chi^2$ or by the routine not converging to a solution.

Our approach is more similar to the ``optimal sampling'' technique introduced by \citet{SPK} to solve the opposite problem: instead of deriving the parameters $\{\alpha,x_m,x_M\}$ from the observations, the authors were interested in generating ``optimally sampled'' simulated data starting from a given set $\{\alpha,x_m,x_M\}$ and an assumed distribution function. Rather than randomly sampling the continuous parental distribution to create a set of discrete simulated data, \citet{SPK} divided the interval between $x_m$ and $x_M$, in our notation, into $N-1$ segments under the condition $\int_{\xi_i}^{\xi_{i+1}}Np(x)\text{d}x=1$, which is exactly the scheme we adopted for equation~(\ref{equPro}). In this way, \citet{SPK} demonstrated that the simulated data reproduce precisely the number distribution, $\text{d}n/\text{d}x$, as well as the distribution of the observable, say the mass, $M\text{d}n/\text{d}x$ of the parental distribution.

Our assumption on the uniformity of the data distribution, so that we construct intervals of equal probability, can look arbitrary. But one should not forget that any solution based on MLE relies on the assumption that the true $\alpha$ maximizes the likelihood to have the observed set $\{x_i\}$. Also this assumption contains a certain level of arbitrariness, even if well reasonable, because it cannot be taken for granted that the intrinsic $\alpha$ generates the most probable set $\{x_i\}$ every time, otherwise randomness would not exist. But it is reasonable to assume that, most frequently, the most probable set will be, indeed, generated. In the same way, we can assume that uniformity of the data distribution, most frequently, will be observed. This assumption is somehow enforced by the work of \citet{SPK} quoted before. Their result shows that the expectation values given by equation~(\ref{expV}) represent the optimal way to create a simulated set of data; our method looks for the power law that, if sampled according to \citet{SPK}'s recipe, would generate expected values as close as possible to the observed set of data.

We also note that the physical mechanism(s) acting on the observed system might generate power-law distributions that are not completely random, for instance causing clustering of data in particular bins. In this case our method will fail to find the correct solution, probably computing a high $\chi^2$ in correspondence of the true $\alpha$. On the other hand, it is likely that any method that assumes perfect randomness in the data distribution, including MLEs, will fail. A correct $\alpha$ will be found only if the physics of the system is known and is inserted when modelling the expected data distribution.


\subsection{The maximum likelihood estimator}
\subsubsection{Finding the exponent $\alpha$}\label{HillE}
For the Pareto distribution with $x_M\rightarrow\infty$, the maximum likelihood estimator (MLE) is \citep[e.g.,][]{Clauset,Newman}
\begin{equation}
\alpha_\text{S}=\frac{N}{\sum\limits_{i=1}^N\ln\frac{x_i}{x_m}}+1\,\,\alpha>1\label{aM1}
\end{equation}
where S stands for Standard, as opposite to Truncated, see below.


The MLE $\alpha_\text{S}$ is sometimes referred to as Hill's estimator after \citet{Hill} who first derived an estimator for $\alpha$ \citep[see, e.g.,][]{brilhante,Huang,nuyts,Cohen,smith}\footnote{For a bibliography up to 2001, see \citet{vollmer}.}. We note, however, that Hill solved a more general problem, different from that expressed with equation~(\ref{px}). The following text extracted from \citet{Hill}'s paper clarifies the difference ($G$ is the distribution followed by the observable $y$): ``On the basis of theoretical arguments or previous data it is believed, or at least the hypothesis is tentatively entertained, that $G$ has a known functional form, say, $G(y)=w(y;\theta)$, \textit{for $y$ sufficiently large}, where $\theta$ is a vector of parameters'' (text emphasized by Hill).

In this approach, the distribution $G$ is thought to describe the entire data set and only in a restricted interval, the tail of the distribution, we assume a given functional form. The tail starts, or ends, in correspondence of an unknown integer $r\le N$: the first, or the last, $r$ values follow the known distribution, a power law in our case, while the remaining $N-r$ do not. The main limitation in using Hill's estimator is indeed the difficulty in deriving $r$ \citep{ResSt}. 

As a consequence, equation~(\ref{CM1}) does not hold because the integral of $G$ can be normalized to 1, while $\int p(y)\text{d}y<1$ can not: its value is unknown and smaller than 1. $C$ is one element of the vector $\theta$ that must be found along with $\alpha$. Hill's estimator $\alpha_\text{H}$ can be written as
\begin{numcases}{\alpha_\text{H}=}
+\frac{N}{\sum\limits_{i=1}^N\ln\frac{x_i}{x_m}}&$\alpha>+1;\,\,x_m\le x<\infty$\label{HM1}
\\
-\frac{N}{\sum\limits_{i=1}^N\ln\frac{x_M}{x_i}}&$\alpha<-1;\,\,0<x\le x_M$.\label{HM2}
\end{numcases}
\citep[e.g.,][]{vollmer}. In the interval $-1\le\alpha\le+1$, $p(x)$ can be normalized only when the distribution is truncated.

Note that in this approach, $x_m$ and $x_M$ may lack a precise definition because $G$ tends to $w$ smoothly and only asymptotically we can say that our data follow the power-law distribution \citep[which explains why it is difficult to derive $r$, see][]{smith}.

When the distribution is truncated, $x_m\le x<x_M$, $C$ is given by equation~(\ref{defC}) and the MLE $\alpha_\text{T}$ is found through the numerical solution of the following equation \citep[][for the discrete case]{aban,Bauke2007}
\begin{equation}
\frac{N}{\alpha_\text{T}-1}+\frac{N\ln f}{1-f^{\alpha_\text{T}-1}}=\sum\limits_{i=1}^N\ln\frac{x_i}{x_m}\label{aTr}
\end{equation}
where, as before, $f=x_M/x_m$. The same equation is found also in \citet{Cohen}, where the ratio $x_m/x_M$ is used and the exponent is $\alpha_\text{T}-1$. Equation~(\ref{aM1}) is recovered in the limit $x_M\rightarrow\infty$, as shown in Appendix~\ref{forF}. To find $\alpha_\text{T}$, we need now not only an estimate of $x_m$ but also of $x_M$.

\subsubsection{Choice of boundary values $x_m$ and $x_M$}\label{sceltaX}
To use equation~(\ref{aM1}) we need an estimate of $x_m$. For equation~(\ref{aTr}) we need an estimate of $x_M$ too. \citet{aban} showed that, asymptotically, the minimum and the maximum of the observed data are indeed equal to $x_m$ and $x_M$. Asymptotically means that the number $N$ of samples must be high enough that we can consider $N\rightarrow\infty$. On the other hand, in general, $N$ can be quite small so that it is worth deriving the error associated to wrong choice of these two parameters.

Consider equation~(\ref{aM1}) and suppose that a wrong value, say $z$, is used as an estimate of $x_m$. In Appendix~\ref{wXM} the following relation is derived, where $\bar\alpha$ is the MLE computed with $z$ instead of $x_m$
\begin{equation}
\bar\alpha=1+\left[\frac1{\alpha_\text{S}-1}+\ln\frac{x_m}z\right]^{-1}\label{zm}.
\end{equation}
This relation does not depend on the size $N$ of the sample but only on the $x_m/z$ ratio. To see quantitatively how $\bar\alpha$ and $\alpha_\text{S}$ are linked, equation~(\ref{zm}) can be written as
\begin{equation}\label{barAAS}
\frac1{\bar\alpha-1}=\frac1{\alpha_\text{S}-1}+\ln\frac{x_m}z\Rightarrow\systeme{
\frac1{\bar\alpha-1}>\frac1{\alpha_\text{S}-1}\,\,\ln(x_m/z)>0\text{,},
\frac1{\bar\alpha-1}<\frac1{\alpha_\text{S}-1}\,\,\ln(x_m/z)<0.}
\end{equation}

When $z<x_m$ then $\ln(x_m/z)>0$ so that $\alpha_\text{S}-1>\bar\alpha-1$ or $\alpha_\text{S}>\bar\alpha$: our estimate is smaller than the MLE value. On the contrary, if $z>x_m$ then $\ln(x_m/z)<0$, thus $\alpha_\text{S}<\bar\alpha$: we end up with a larger estimate of $\alpha$.

A hidden dependency on $N$ indeed exists because, as already pointed out by \citet{aban}, if we set $z=\text{min}(x_i)$ then $z\rightarrow x_m$ as $N\rightarrow\infty$. In any case, both $\alpha_\text{S}$ and $\bar\alpha$ cannot be smaller than, or equal to 1 (unless one makes the strange assumption $z>\text{min}(x_i)$).

Another problem with equation~(\ref{aM1}), beside the choice of $x_m$ , is that the distribution does not have a limiting value $x_M$ (remember that $x_m\le x<\infty$). As a consequence, a simulated set of data can have extremely high values of $x$. In the simulations discussed later on, in 872 series out of 1,000 we found $\text{max}(x)>500$ for $\alpha=1.5$ and $x_m=0.8$, already after generating 50 data. For longer series we can find even more extreme values like $10^{13}$. Such extreme dynamical ranges are not possible in reality in many astrophysical problems. Thus, these simulation are not representative of a realistic distribution.

We will come again on this point in Section~\ref{simOSB}; here we note that the knowledge of $x_M$ is very relevant from a physical point of view, for instance in the case of CMF: in some molecular clouds we observe the formation of low-mass stars only. Assuming $\sim10\,M_\odot$ as the limit between low- and high-mass stars (the reason for this limit linked to the birth of stars can be found in \citealt{PeS}; or in \citealt{ZeY} for a link to the death of stars) and a star formation efficiency per single core of $\sim0.3$ \citep[e.g.,][]{perseo} the power law describing the CMF must be truncated at $m\sim30\,M_\odot$ to make it correctly reproduce reality\footnote{For the purposes of this paper it is not important to detail the star-formation process, so that we can neglect the role of core fragmentation in shaping the CMF. Taking into account fragmentation would change the value of $x_M$, leaving anyway the necessity to bound the distribution on the high values.}. The IMF itself is limited to a few $10^2\,M_\odot$ \citep{WK2004,Figer2005,BKO2012}.

If we assume that $x_M$ is not finite when it actually is, $\alpha_\text{S}$ will not converge to $\alpha$ even if $N\rightarrow\infty$. To see why, one can compute $\alpha_\text{S}$ with equation~(\ref{aM1}) and $\alpha_\text{T}$ with equation~(\ref{aTr}). As $\sum_{i=1}^N\ln(x_i)$, as well as the estimate of $x_m$, are the same in both equations, the following relation between the two exponents can be derived
\begin{equation}\label{alphaSalphaT}
\alpha_\text{S}=1+\left[\frac1{\alpha_\text{T}-1}+\frac{\ln f}{1-f^{\alpha_\text{T}-1}}\right]^{-1}.
\end{equation}

Now, $\alpha_\text{S}$ and $\alpha_\text{T}$ are the same as long as $f\rightarrow\infty$. But if $f$ is small, we can see the consequence in equation~(\ref{alphaSalphaT}) by noticing that when $f\rightarrow1$, then $\alpha_\text{S}\rightarrow\infty$ (see Appendix~\ref{forF}). In other words, if our data follow a truncated power-law distribution with exponent $\alpha$, the estimate $\alpha_\text{S}$ gets worse and worse as $f\rightarrow1$.

\section{Simulations}\label{secSim}
Equations~(\ref{aM1}) and (\ref{aTr}) are already known, so there is no need to check their reliability for large samples. On the other hand, it is interesting to see how well the estimators of $\alpha_\text{S}$ behave when the factor $f$ is not infinite, as assumed when deriving equation~(\ref{aM1}); also, we want to check the reliability of these solutions when used with small samples of data. In fact, we should not forget that the MLE technique is valid only asymptotically for large samples \citep[][]{Clauset,ago}. Instead, our method is new, so we need to verify the correctness of the solution for small samples and when $N\rightarrow\infty$.

To simulate power-law distributions, we have generated 1,000 series of random numbers following a uniform distribution, each series consisting of 10,000 values. With this set of data we build all the simulated power-law distributions, as explained in Appendix~\ref{genera}. To create one-side-bounded simulated data ($x_m\le x<\infty$) equation~(\ref{xFr}) is used, with $\alpha$ in $\{1.5, 2.0, 2.5, 3.5\}$. The first three values are chosen because our interest is in the CMF for which we expect values in this range. The highest value is adopted because $\alpha>2.5$ can be relevant for CMF and IMF too (references in Section~\ref{CMFP}). For $x_m$ we follow \citet{maizUbe} fixing the value to 0.8; note, however, that the precise value of $x_m$ is not important, because equation~(\ref{aM1}) is scale-free, it does not change if all the simulated data are multiplied by a scale factor.

The length of the series, 10,000 points, is chosen to make sure that all the algorithms reach the asymptotic value for the parameter(s): this is confirmed by comparing the results found considering the first 1,000 points with those found using 5,000 points and the entire series of 10,000 values. The number of simulations, 1,000, should be high enough to ensure that the standard deviation of the mean of the parameter(s) is accurately measured. As a metric to judge the performance of the different methods we adopt a 10\% criterion: the true value $q$ is recovered when $\bar{p}_n\pm1\sigma\le0.1q$, $\bar{p}_n$ being the value of one of the parameters averaged over all the simulations using the first $n$ values. We set the minimum number of points to 4, because this is required by the least-squares fitting routine, so that we have $p_4$, $p_5$ $\ldots$ up to $p_{10,000}$.

For the two-sides-bounded distributions ($x_m\le x<x_M$) we start from the same set of 1,000 simulations, but now using equation~(\ref{xTFr}) to create the simulated data. The same set of $\alpha$ and same $x_m$ used for the one-side-bounded case are adopted, while for the upper limit we set $x_M=40$. Actually, as we shall see, what is important is $f=x_M/x_m$ rather than $x_M$ itself: the value $f=50$ is small enough to see already the effect on $\alpha_\text{S}$.

In the following, SMLE stands for standard MLE whose exponent, $\alpha_\text{S}$, is found using equation~(\ref{aM1}); TMLE stands for truncated MLE where the slope, $\alpha_\text{T}$, is derived through equation~(\ref{aTr}); LST is used for our least-squares method whose exponent, $\alpha_\text{L}$, is derived with equations~(\ref{lS}). As written, the 1,000 series of uniformly distributed simulated data are available on-line (URL given before References).

\subsection{The case of one-side-bounded distributions}\label{simOSB}
The convergence speed of each technique, as said, is measured with a 10\% criterion. In Table~\ref{osb} we report the minimum number of points necessary to meet this criterion: for the slope, this number increases with $\alpha$.

\begin{table}
	\centering
	\caption{Minimum number of points needed to have $\bar\alpha\pm1\sigma\le0.1\alpha$, where $\alpha$ is the true value, written in the first row, and $\bar\alpha$ is the mean of the 1,000 simulations for SMLE ($\alpha_\text{S}$), TMLE ($\alpha_\text{T}$), and LST ($\alpha_\text{L}$), respectively. For $x_m$ the criterion is the same, i.e., $\bar{x}_m\pm1\sigma\le0.1x_m$. This parameter can be computed only with our least-squares method.}
	\label{osb}
	\begin{tabular}{lcccc|lcccc} 
		\hline
		& 1.5  & 2.0 & 2.5 & 3.5 & & 1.5  & 2.0 & 2.5 & 3.5\\
		\hline
		$\alpha_\text{S}$ & 7 & 12 & 19 & 29 & $\alpha_\text{L}$ & 25 & 34 & 41 & 53\\
		$\alpha_\text{T}$ & 10 & 19 & 26 & 33 & $x_m$ & 136 & 20 & 9 & 5\\
		\hline
	\end{tabular}
\end{table}

\begin{figure}
	\includegraphics[width=\columnwidth]{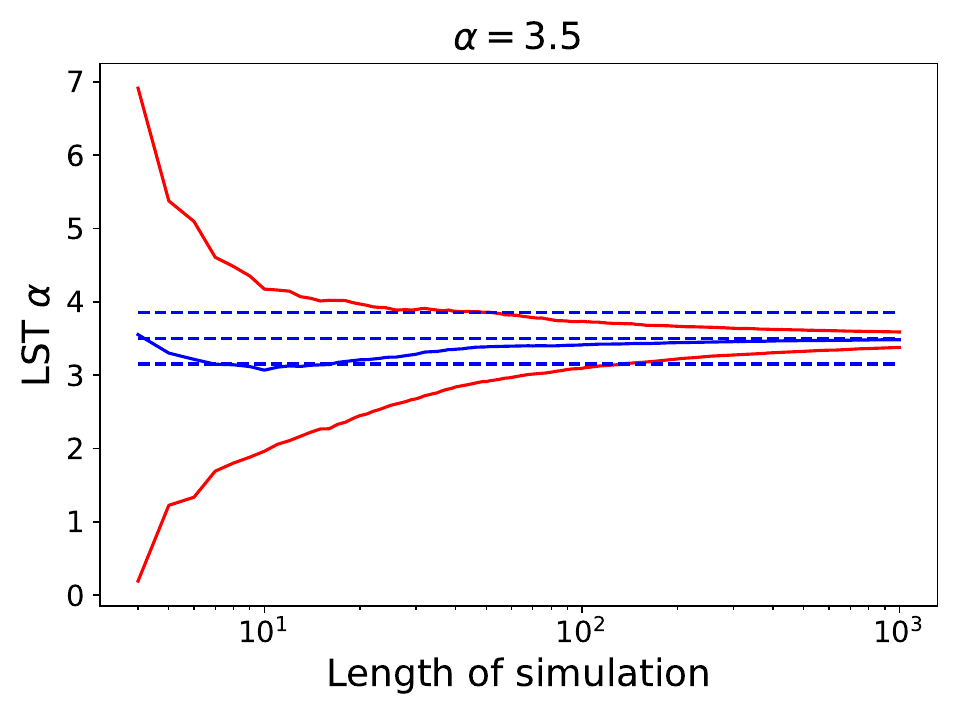}
	\includegraphics[width=\columnwidth]{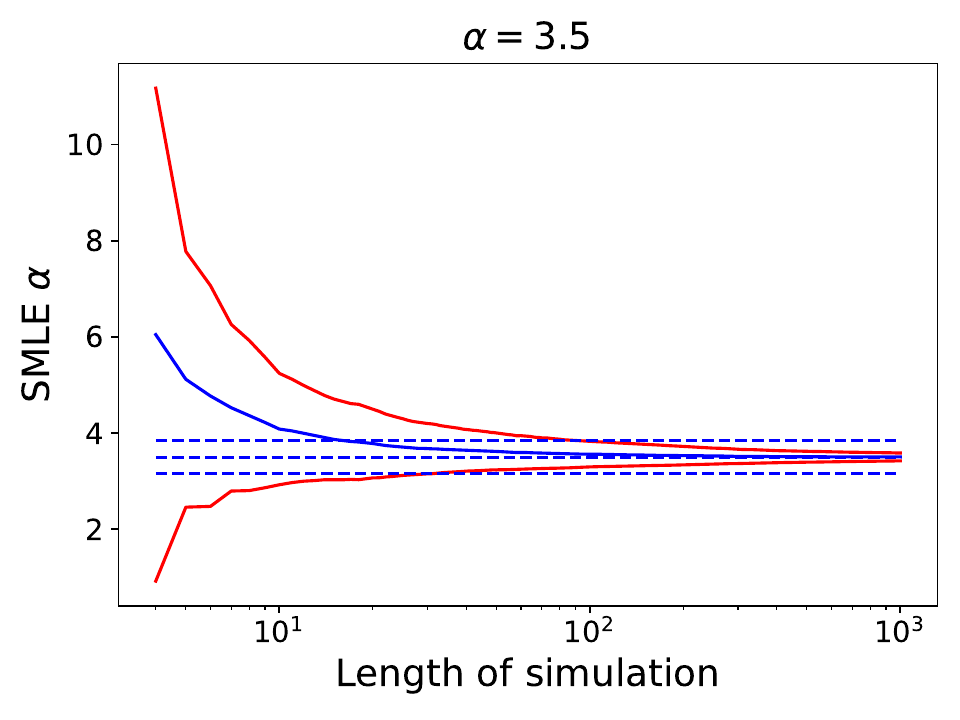}
	\includegraphics[width=\columnwidth]{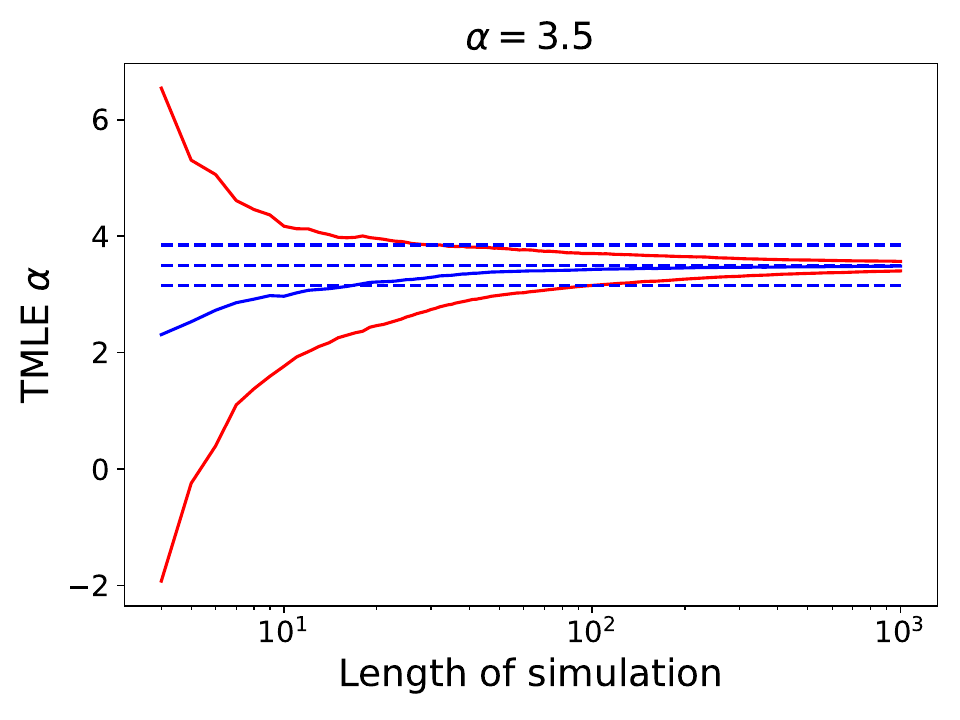}
    \caption{Derivation of the slope for 1,000 simulated power-law distributions defined by equation~(\ref{pxnt}). In the panels, from top to bottom, the exponents $\alpha$ were computed through equation~(\ref{lS}), equation~(\ref{aM1}), and equation~(\ref{aTr}), respectively. Simulated data were obtained with $x_m=0.8$ and $\alpha=3.5$. In all panels, x-axis is the length of the simulations, from 4 to 1,000. Solid blue line: the computed slope averaged over the 1,000 simulations; dashed blue lines: true value and 10\% intervals; red lines: $\bar\alpha\pm1\sigma$.}
    \label{osb3.5}
\end{figure}

Being $\alpha=3.5$ the worst case, in Fig.~\ref{osb3.5} we show for this exponent the trend of $\bar\alpha$ with $N$, averaging for each $N$ in the range 4 -- 1,000, the values of the 1,000 slopes, one for each simulation. The results for the other $\alpha$'s are qualitatively similar. The mean of the 1,000 computed $\bar\alpha_\text{L}$ (blue line) using equation~(\ref{lS}) are shown in the top panel; means $\bar\alpha_\text{S}$ computed with equation~(\ref{aM1}) are in the central panel; finally, the results for $\bar\alpha_\text{T}$ derived through equation~(\ref{aTr}) are in the bottom panel.

The three methods give results which are not much different to each other. Note that using equation~(\ref{aM1}), $\alpha_\text{S}$ is always larger than $\alpha$, an effect predicted by equation~(\ref{barAAS}): in fact, $x_m$ is estimated as $\text{min}(x_i)$ but clearly $\text{min}(x_i)\ge x_m$ or, in the formalism of equation~(\ref{barAAS}), $z\ge x_m$. On the contrary, $\alpha_\text{T}$ is always smaller than $\alpha$ when using equation~(\ref{aTr}). The reason why $\alpha_\text{T}<\alpha$ when, in equation~(\ref{aTr}), we set $x_m=\text{min}(x_i)$ and $x_M=\text{max}(x_i)$, is not easy to understand. What we can say is that, for $1<f<\infty$, from equation~(\ref{alphaSalphaT}) we have $\infty>\alpha_\text{S}/\alpha_\text{T}>1$. And since $f$ is estimated from the simulated data as $\text{max}(x_i)/\text{min}(x_i)<\infty$, then, for any finite value of $f$ we have $\alpha_\text{S}>\alpha_\text{T}$.

The numerical results are summarized in Table~\ref{msObs} where we report the numerical values of means and standard deviations of the parameter(s) found with the different methods, for small and large $N$.

\begin{table*}
	\centering
	\caption{Mean and standard deviations, averaged over the 1,000 simulations, of the parameters $\alpha_\text{L}$, $x_m$ and $x_M$ found with our technique, equation~(\ref{lS}), and of $\alpha_\text{S}$ found with SMLE, equation~(\ref{aM1}) and of $\alpha_\text{T}$ found with TMLE, equation~(\ref{aTr}), using the number of points given in column $N$. The case $N=10,000$ corresponds to the entire series. Simulated data are in the range $0.8\le x<\infty$.}
	\label{msObs}
	\begin{tabular}{rccc|ccccc} 
		\hline
		$N$&$\alpha_\text{L},x_m,x_M$&$\alpha_\text{S}$&$\alpha_\text{T}$&$\alpha_\text{L},x_m,x_M$&$\alpha_\text{S}$&$\alpha_\text{T}$\\
		\hline
		&\multicolumn{3}{c}{$\alpha=3.5$,$x_m=0.8$, $x_M$ not constrained}&\multicolumn{3}{c}{$\alpha=2.0$,$x_m=0.8$, $x_M$ not constrained}\\
		10&$3.0\pm1.1$      &$4.0\pm1.2$    &$2.9\pm1.2$  &$1.87\pm0.45$&$2.23\pm0.46$&$1.78\pm0.48$\\
		  &$0.785\pm0.056$&&&$0.79\pm0.12$&&\\
		  &$3.7\pm3.2$&&&$80\pm550$&&\\
       100&$3.41\pm0.32$    &$3.56\pm0.27$  &$3.42\pm0.27$&$1.98\pm0.13$&$2.02\pm0.11$&$1.97\pm0.11$\\
          &$0.793\pm0.022$&&&$0.791\pm0.053$&&\\
          &$8.8\pm8.9$&&&$1,000\pm13,000$&&\\
     1,000&$3.48\pm0.11$    &$3.503\pm0.082$&$3.486\pm0.082$&$1.998\pm0.043$&$2.001\pm0.036$&$1.994\pm0.033$\\
          &$0.7986\pm0.0086$&&&$0.799\pm0.021$&&\\
          &$20\pm15$&&&$4,000\pm19,000$&&\\
     5,000&$3.494\pm0.047$  &$3.501\pm0.035$&$3.497\pm0.035$&$1.999\pm0.019$&$2.000\pm0.014$&$1.998\pm0.014$\\
          &$0.7994\pm0.0042$&&&$0.799\pm0.010$&&\\
          &$42\pm42$&&&$>4\cdot10^4$&&\\
    10,000&$3.496\pm0.036$  &$3.500\pm0.025$&$3.498\pm0.025$&$1.999\pm0.015$&$2.000\pm0.010$&$1.999\pm0.010$\\
          &$0.7995\pm0.031$&&&$0.7998\pm0.0079$&&\\
          &$56\pm49$&&&$>7\cdot10^4$&&\\\hline
		&\multicolumn{3}{c}{$\alpha=2.5$,$x_m=0.8$, $x_M$ not constrained}&\multicolumn{3}{c}{$\alpha=1.5$,$x_m=0.8$, $x_M$ not constrained}\\
        10&$2.25\pm0.66$    &$2.84\pm0.69$  &$2.18\pm0.72$&$1.50\pm0.25$&$1.61\pm0.23$&$1.39\pm0.24$\\
          &$0.784\pm0.086$&&&$0.88\pm0.25$&\\
          &$12\pm29$&&&$>3\cdot10^5$&&\\
       100&$2.45\pm0.20$    &$2.53\pm0.16$&$2.45\pm0.16$  &$1.503\pm0.069$&$1.512\pm0.053$&$1.485\pm0.055$\\
          &$0.791\pm0.036$&&&$0.802\pm0.093$&\\
          &$50\pm200$&&&$>1\cdot10^8$&&\\
     1,000&$2.492\pm0.063$    &$2.502\pm0.049$&$2.491\pm0.049$&$1.503\pm0.022$&$1.500\pm0.016$&$1.497\pm0.016$\\
          &$0.798\pm0.014$&&&$0.807\pm0.043$&\\
          &$190\pm350$&&&$>4\cdot10^8$&&\\
     5,000&$2.498\pm0.029$  &$2.500\pm0.021$&$2.498\pm0.021$&$1.5016\pm0.0099$&$1.5002\pm0.0071$&$1.4994\pm0.0070$\\
          &$0.7994\pm0.0070$&&&$0.803\pm0.022$&&\\
          &$700\pm2,200$&&&$>10^{11}$&&\\
    10,000&$2.498\pm0.022$  &$2.500\pm0.015$&$2.498\pm0.015$&$1.5010\pm0.0076$&$1.5001\pm0.0051$&$1.4996\pm0.0051$\\
          &$0.7995\pm0.0053$&&&$0.802\pm0.017$&&\\
          &$1,100\pm2,600$&&&$>10^{11}$&&\\\hline
	\end{tabular}
\end{table*}


Being not constrained, it is not surprising that $x_M$ grows without limits as $N$ increases. However, it should be noted that the higher $\alpha$, the smaller is $x_M$: this is because a steep power law will require more and more simulated data to generate high values of the observable. In fact, for $\alpha=3.5$, among the 1,000 simulations and considering the entire series of 10,000 values, the largest $\text{max}(x_i)$ is only 1,424, while it can be as low as 14. This consideration will play an important role when dealing with two-side-bounded distributions.

\begin{figure}
 	\includegraphics[width=\columnwidth]{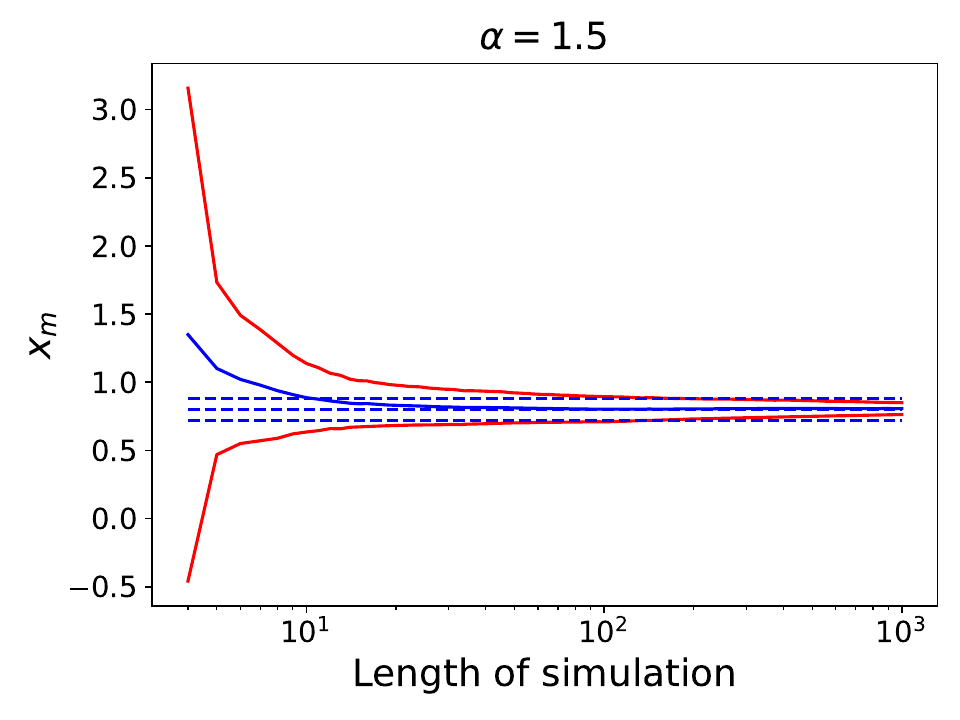}
     \caption{The derived value of $x_m$ for $\alpha=1.5$. Lines and colors as in Fig.~\ref{osb3.5}.}
     \label{x11.5}
 \end{figure}

For $x_m$, $N$ increases when $\alpha$ decreases. As visible in Fig.~\ref{x11.5}, the reason for this slow convergence is due to the large scatter of the 1,000 $x_m$’s derived for each simulation. The average of these values needs less than 10 data to give the correct answer within 10\%; $\bar{x}_m-\sigma_{x_m}$ recovers the input value with less than 20 data within 15\%, still a valuable result.

 
\subsection{The case of two-sides-bounded distributions}\label{simOTB}
In Table~\ref{otb} we show, as before, the minimum number $N$ of data such that, on average, the mean of each parameter differ, from the true value, by less than 10\% within $1\sigma$. When dealing with two-sides-bounded distributions, however, the interpretation of this table is not immediate. In particular, the trend of $N$ with $\alpha$ for $\alpha_\text{S}$ is odd: $N$ decreases with $\alpha$ down to 2.0, then, for $\alpha=1.5$, equation~(\ref{aM1}) never converges to the expected result, not even using the entire series ($N=10,000$).

\begin{table}
	\centering
 	\caption{Same as Table~\ref{osb} for the case of truncated power-law distributions. Parameters $x_m$ and $x_M$ are found only with our least-squares method.}
 	\label{otb}
	\begin{tabular}{lcccc|lcccc} 
		\hline
		& 1.5  & 2.0 & 2.5 & 3.5 & & 1.5  & 2.0 & 2.5 & 3.5\\
		\hline
		$\alpha_\text{S}$ & -- & 13 & 19 & 29 & $\alpha_\text{L}$ & 66 & 54 & 48 & 53\\
		$\alpha_\text{T}$ & 39 & 19 & 24 & 33 & $x_m$ & 50 & 19 & 9 & 5\\
		&&&&& $x_M$ & 8 & 36 & 171 &$>5\cdot10^3$\\
		\hline
	\end{tabular}
\end{table}

\begin{figure}
	\includegraphics[width=\columnwidth]{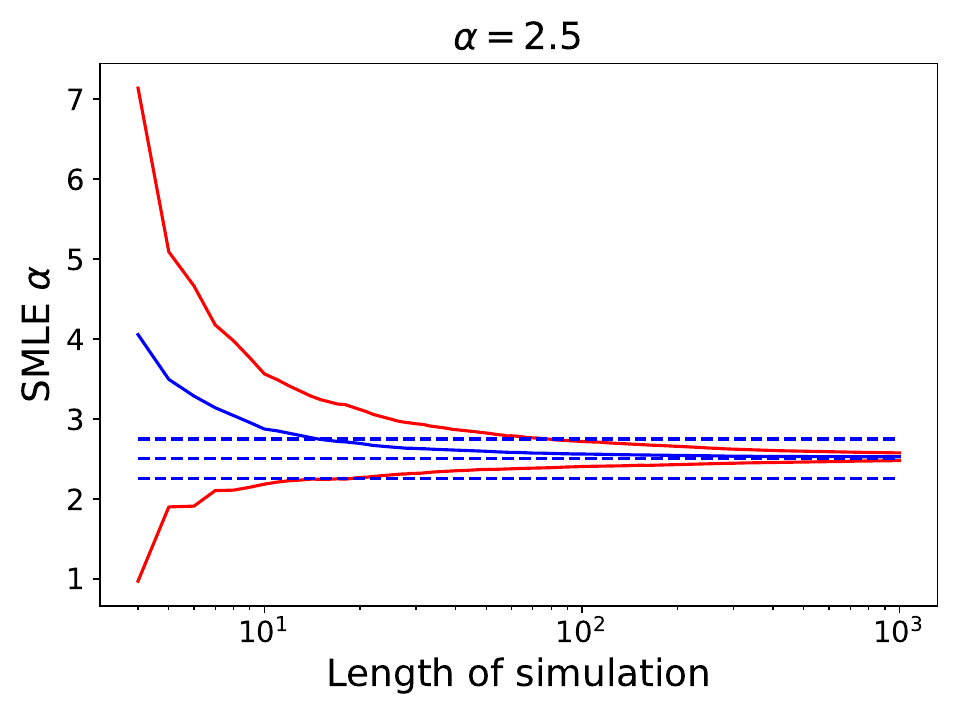}
	\includegraphics[width=\columnwidth]{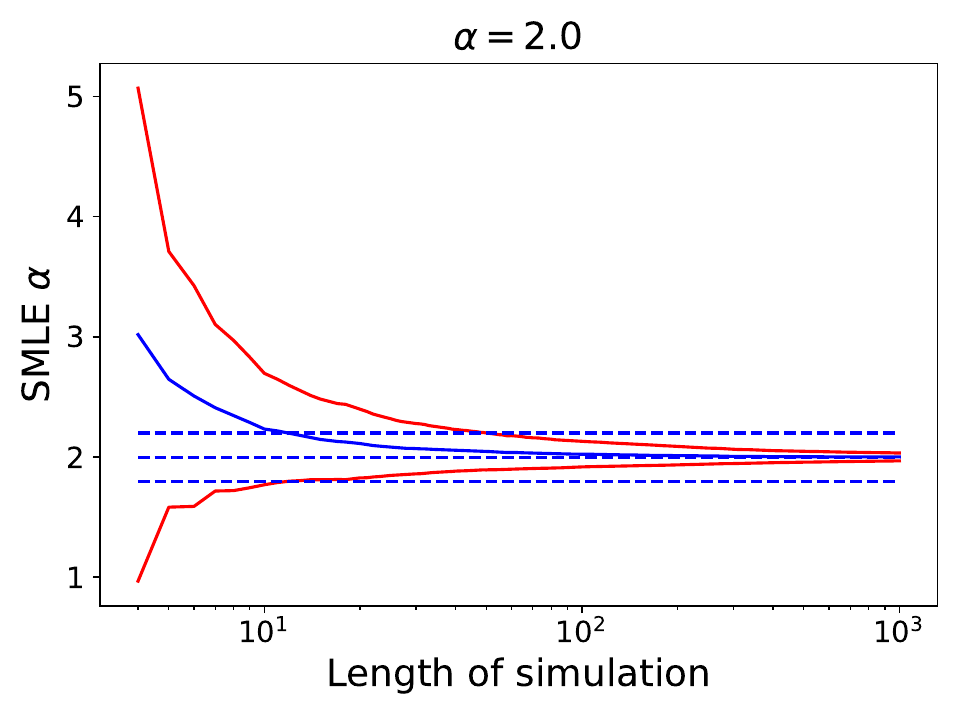}
	\includegraphics[width=\columnwidth]{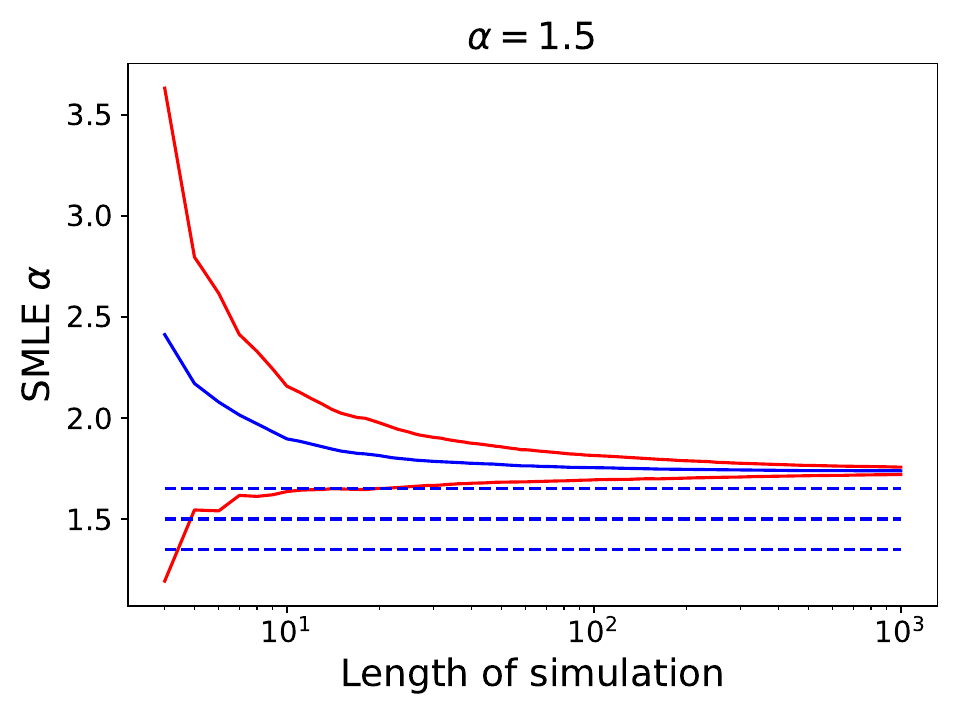}
    \caption{Derivation of the slope for 1,000 simulated power-law distributions defined by equation~(\ref{pxt}) with $x_m=0.8$ and $x_M=40$, found with equation~(\ref{aM1}), and for three different values of $\alpha$ shown at the top of each panel. Lines and colors as in Fig.~\ref{osb3.5}.}
    \label{SMLESba}
\end{figure}

To understand this behaviour, in Fig.~\ref{SMLESba} we show $\alpha_\text{S}$ vs. $N$ for $\alpha=2.5$, 2.0 and 1.5. While for $\alpha=2.5$ and 3.5 (not shown) equation~(\ref{aM1}) works well, for smaller $\alpha$ the fact that $f$ is finite, $f=50$ in our simulations, starts playing a role and equation~(\ref{aM1}), derived under the assumption that $f$ is infinite, does not provide the correct answer. For $\alpha=2.0$ the effect is still small, only 7 values are enough to have $\bar\alpha_\text{S}\pm1\sigma_\text{S}\le0.1\alpha$, but the asymptotic value of $\alpha_\text{S}$ is not correct. In other words, the standard MLE is not consistent: for $N\rightarrow\infty$, $\alpha_\text{S}$ does not converge to $\alpha$. The effect becomes much more evident for $\alpha=1.5$.

\begin{figure}
	\includegraphics[width=\columnwidth]{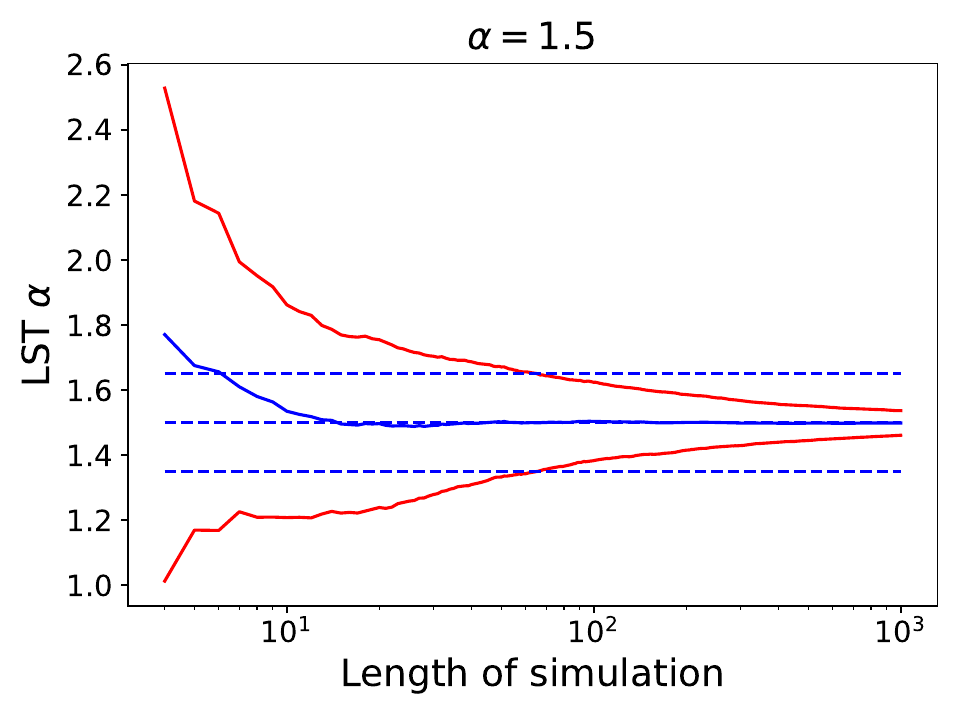}
	\includegraphics[width=\columnwidth]{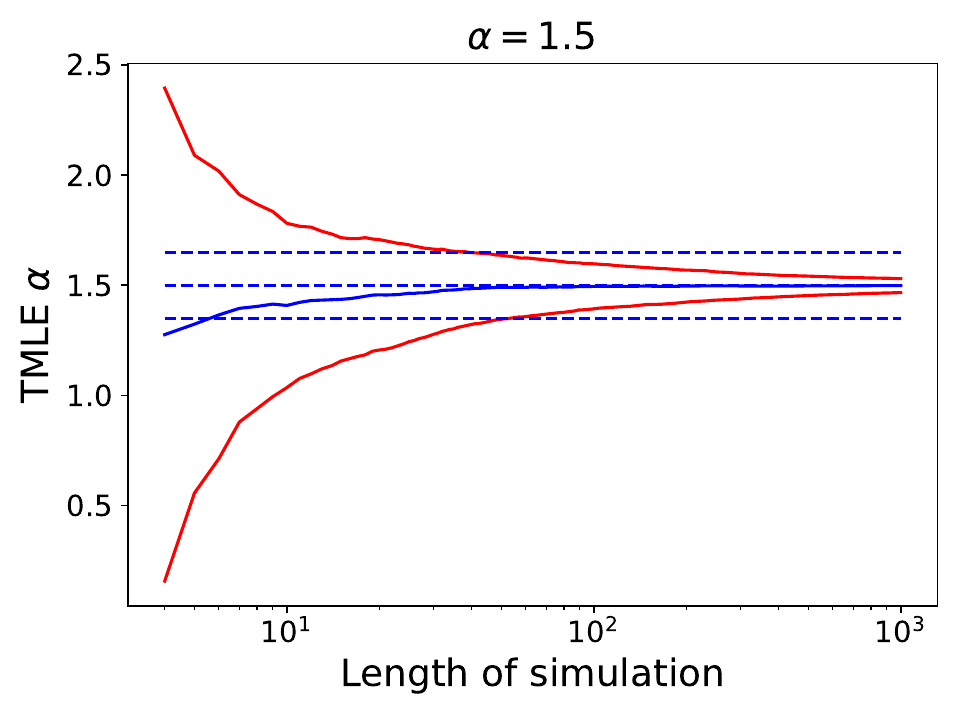}
    \caption{Derivation of the slope for 1,000 simulated power-law distributions defined by equation~(\ref{pxt}) with $\alpha=1.5$, $x_m=0.8$ and $x_M=40$, found with equation~(\ref{lS}), top panel, and equation~(\ref{aTr}), bottom panel. Lines and colors as in Fig.~\ref{osb3.5}.}
    \label{buone1.5}
\end{figure}

On the contrary, both $\alpha_\text{T}$, derived from equation~(\ref{aTr}), and $\alpha_\text{L}$, derived with our method, converge to the correct result. There is no trend of $N$ with $\alpha$, see Table~\ref{otb}, but we can conclude that when $N\ga40$, both methods provide good estimates of $\alpha$. The worst case is for $\alpha=1.5$ as we show in Fig.~\ref{buone1.5}. This value of $\alpha$ makes it possible also to compare $\alpha_\text{L}$ and $\alpha_\text{T}$ with $\alpha_\text{S}$ (bottom panel of Fig.~\ref{SMLESba}).

As equations~(\ref{aM1}) and (\ref{aTr}) differ by the factor $f$, we can give a rough estimate of when equation~(\ref{aM1}) starts giving wrong results. Considering our set of $\alpha$, $f^{1-\alpha}$ takes on the values $5.7\times10^{-5}$, $2.8\times10^{-3}$, 0.02 and 0.14 for $\alpha=3.5$, 2.5, 2.0 and 1.5, respectively. Our results, based however on a small sample of the parameter space, suggest that for $\alpha_\text{S}=2.0$ the SMLE starts being not correct; thus, we tentatively suggest that using equation~(\ref{aM1}) for $f^{1-\alpha}\ga0.02$ can lead to potentially wrong results ($\alpha_\text{S}>\alpha$). For instance, the asymptotic value of $\alpha_\text{S}$ is 1.7378 when $\alpha=1.5$, and since $f=50$ can be easily estimated from the data, one finds $f^{1-\alpha_\text{S}}=0.056>0.02$ which means that our value for $\alpha_\text{S}$ is potentially wrong.

\begin{figure}
	\includegraphics[width=\columnwidth]{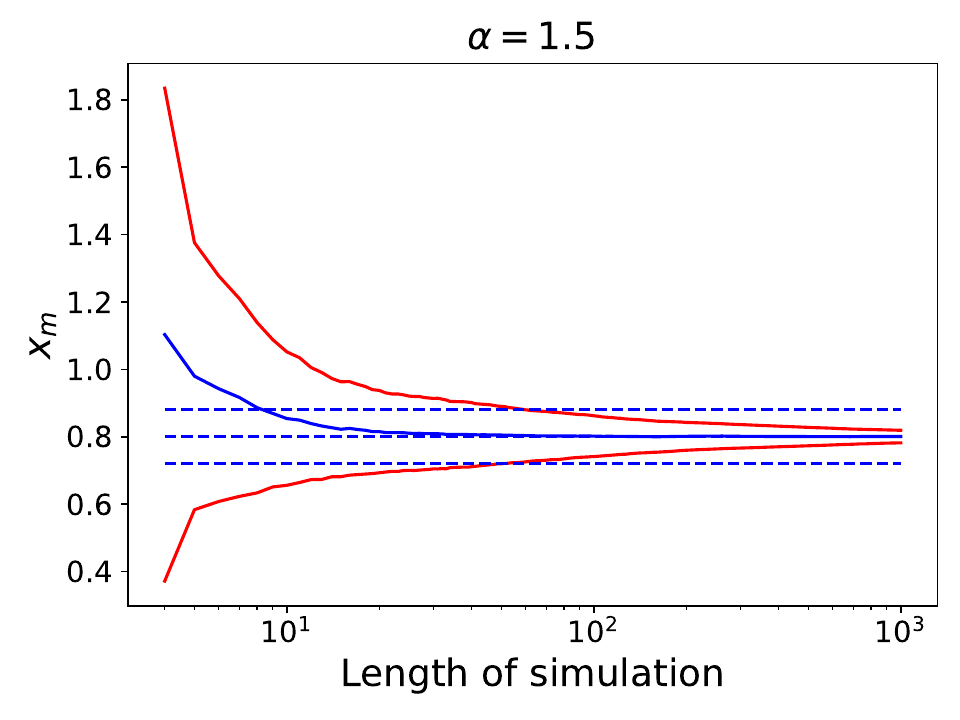}
    \caption{Derivation of $x_m$ through equation~(\ref{lS}) when $\alpha=1.5$. Lines and colors as in Fig.~\ref{osb3.5}.}
    \label{x11.5bis}
\end{figure}

The trend of $N$ with $\alpha$ for $x_m$ shows the same behaviour of the single-bound distributions case. Also for the truncated distribution the average value $\bar{x}_m$ needs less than 10 data to give the correct answer within 10\% and $\bar{x}_m-\sigma_{x_m}$ recovers the input value with $N\sim10$ within 15\%. The worst case ($\alpha=1.5$) is shown in Fig.~\ref{x11.5bis}.

\begin{figure}
	\includegraphics[width=\columnwidth]{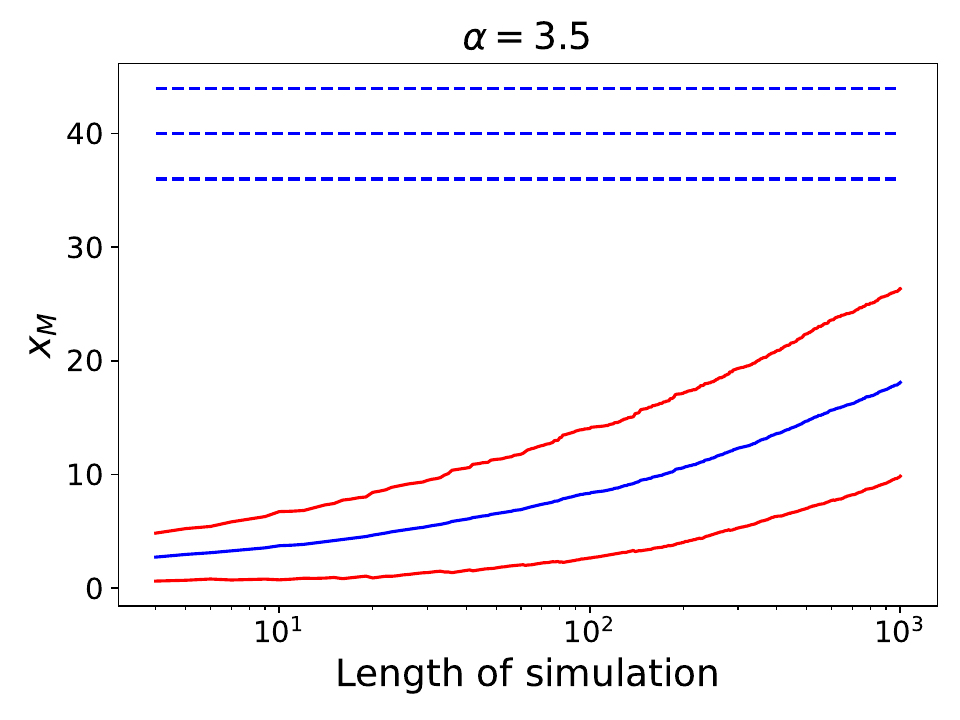}
    \caption{Derivation of $x_M$ through equation~(\ref{lS}) for $\alpha=3.5$. Lines and colors as in Fig.~\ref{osb3.5}.}
    \label{x23.5}
\end{figure}

The results for $x_M$ are interesting. From Table~\ref{otb} we can see that $x_M$ needs long series to match our 10\% criterion when $\alpha=3.5$. Also for smaller $\alpha$ the convergence is very slow; only when $\alpha=1.5$ the criterion is met with a reasonable length of the simulations. The reason for this, however, is not due to our method. To clarify this point, we show in Fig.~\ref{x23.5} the trend of the mean $x_M$ with $N$ in the worst case of $\alpha=3.5$. When $N=1,000$ the mean of $x_M$ is $18.1\pm8.3$ (see Table~\ref{msTr}).

\begin{figure}
	\includegraphics[width=\columnwidth]{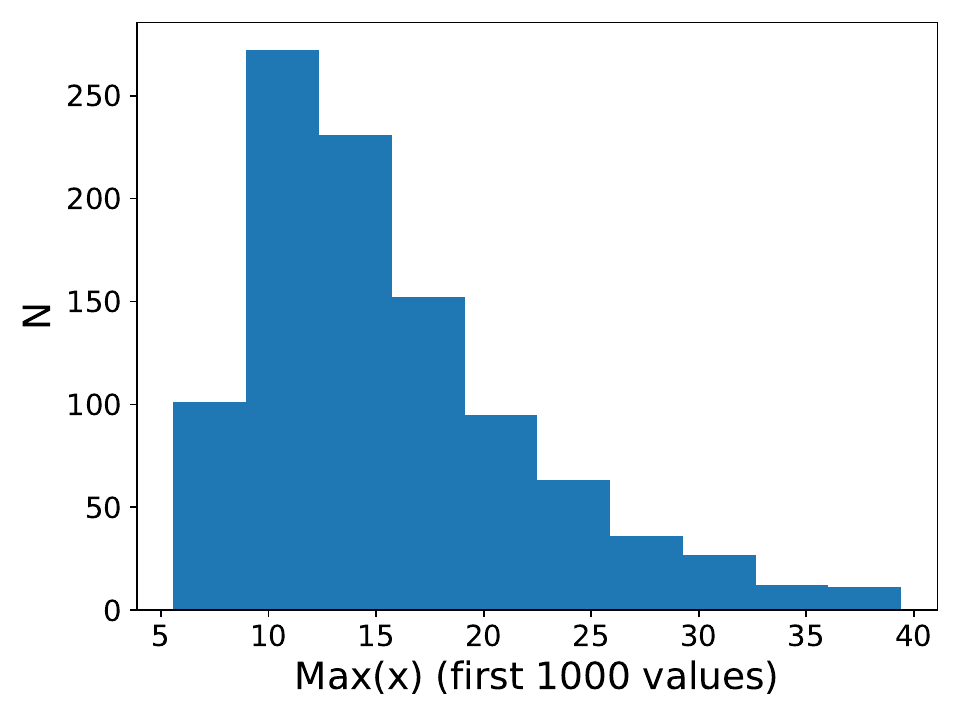}
    \caption{Histogram of the maximum of each simulation, considering the first 1,000 values of each series, when $\alpha=3.5$.}
    \label{distro}
\end{figure}

The reason for that is the steepness of the power law. This is visible in Fig.~\ref{distro} where we show the distribution of the maximum of the first 1,000 simulated values of each simulation, when $\alpha=3.5$. The distribution peaks at very small values, with about 50\% of the simulations having the maximum between $\sim9$ and $\sim16$. Under these conditions, the derived value of $\bar{x}_M\sim18$ can, indeed, be considered a correct result. For comparison, the 1,000 maxima of the simulations for $\alpha=1.5$, again considering only the first 1,000 points of each simulation, are above 39.6. That is why the convergence is faster.

All the results presented in this section are summarized in Table~\ref{msTr}, which has the same structure as Table~\ref{msObs}.

\begin{table*}
	\centering
	\caption{Mean and standard deviations, averaged over the 1,000 simulations, of the parameters $\alpha_\text{L}$, $x_m$ and $x_M$ found with our technique, equation~(\ref{lS}), and of $\alpha_\text{S}$ found with SMLE, equation~(\ref{aM1}) and of $\alpha_\text{T}$ found with TMLE, equation~(\ref{aTr}), using the number of points given in column $N$. The case $N=10,000$ corresponds to the entire series. Simulated data are in the range $0.8\le x<40$.}
	\label{msTr}
	\begin{tabular}{rccc|ccccc} 
		\hline
		$N$&$\alpha_\text{L},x_m,x_M$&$\alpha_\text{S}$&$\alpha_\text{T}$&$\alpha_\text{L},x_m,x_M$&$\alpha_\text{S}$&$\alpha_\text{T}$\\
		\hline
		&\multicolumn{3}{c}{$\alpha=3.5$,$x_m=0.8$, $x_M=40$}&\multicolumn{3}{c}{$\alpha=2.0$,$x_m=0.8$, $x_M=40$}\\
		10&$3.0\pm1.1$      &$4.0\pm1.2$    &$2.9\pm1.2$  &$1.88\pm0.47$&$2.23\pm0.46$&$1.78\pm0.48$\\
		  &$0.785\pm0.056$&&&$0.79\pm0.12$&&\\
		  &$3.7\pm3.0$&&&$15\pm11$&&\\
       100&$3.41\pm0.32$    &$3.56\pm0.26$  &$3.42\pm0.27$&$1.98\pm0.15$&$2.02\pm0.11$&$1.97\pm0.11$\\
          &$0.793\pm0.022$&&&$0.795\pm0.045$&&\\
          &$8.3\pm5.7$&&&$31.9\pm9.2$&&\\
     1,000&$3.40\pm0.11$    &$3.504\pm0.082$&$3.486\pm0.082$&$1.998\pm0.047$&$2.001\pm0.033$&$1.994\pm0.033$\\
          &$0.7987\pm0.0085$&&&$0.800\pm0.014$&&\\
          &$18.1\pm8.3$&&&$39.1\pm4.8$&&\\
     5,000&$3.495\pm0.047$  &$3.502\pm0.035$&$3.497\pm0.035$&$2.000\pm0.020$&$2.087\pm0.014$&$1.999\pm0.017$\\
          &$0.7995\pm0.0041$&&&$0.8002\pm0.063$&&\\
          &$28.3\pm9.2$&&&$39.8\pm2.3$&&\\
    10,000&$3.496\pm0.035$  &$3.501\pm0.025$&$3.498\pm0.025$&$2.000\pm0.014$&$2.0870\pm0.0098$&$1.999\pm0.012$\\
          &$0.7996\pm0.0030$&&&$0.07999\pm0.0044$&&\\
          &$32.1\pm8.7$&&&$39.9\pm1.8$&&\\\hline
		&\multicolumn{3}{c}{$\alpha=2.5$,$x_m=0.8$, $x_M=40$}&\multicolumn{3}{c}{$\alpha=1.5$,$x_m=0.8$, $x_M=40$}\\
        10&$2.25\pm0.67$    &$2.87\pm0.69$  &$2.18\pm0.73$&$1.53\pm0.33$&$1.89\pm0.26$&$1.40\pm0.37$\\
          &$0.784\pm0.086$&&&$0.85\pm0.20$&\\
          &$8.7\pm7.9$&&&$25\pm12$&&\\
       100&$2.46\pm0.20$    &$2.56\pm0.16$&$2.46\pm0.17$  &$1.50\pm0.12$&$1.754\pm0.060$&$1.49\pm0.10$\\
          &$0.792\pm0.034$&&&$0.801\pm0.060$&\\
          &$21\pm10$&&&$38.0\pm6.2$&&\\
     1,000&$2.495\pm0.063$    &$2.527\pm0.048$&$2.493\pm0.052$&$1.498\pm0.038$&$1.738\pm0.018$&$1.498\pm0.032$\\
          &$0.799\pm0.012$&&&$0.800\pm0.019$&\\
          &$35.8\pm7.3$&&&$39.7\pm2.6$&&\\
     5,000&$2.499\pm0.027$  &$2.526\pm0.021$&$2.499\pm0.022$&$1.500\pm0.016$&$1.7380\pm0.0079$&$1.499\pm0.014$\\
          &$0.7999\pm0.0054$&&&$0.8004\pm0.0083$&&\\
          &$39.4\pm4.6$&&&$39.9\pm1.2$&&\\
    10,000&$2.499\pm0.019$  &$2.525\pm0.015$&$2.499\pm0.016$&$1.500\pm0.011$&$1.7378\pm0.0056$&$1.4999\pm0.0097$\\
          &$0.7998\pm0.0038$&&&$0.8000\pm0.0059$&&\\
          &$39.7\pm3.8$&&&$39.98\pm0.83$&&\\\hline
	\end{tabular}
\end{table*}

\section{Working with real data}\label{realD}
The fundamental difference between simulations and real data is that, in the former case, we know that all the data follow the same power-law distribution; in the latter case, we do not. And even in case data follow such a distribution, it is not known over which range. Moreover, in case of simulations, computing $x_m$ and $x_M$ is necessary only as a test of the method. Their exact value is of little interest because they cannot differ significantly from $\text{min}(x)$ and $\text{max}(x)$ \citep[at least for large samples,][]{aban}. It is with real data that we need to derive $\alpha$, $x_m$ and $x_M$, which, generally, are not known a priori. A further complication is that data uncertainty must be taken into account.

If we have enough data to build a histogram, its shape can suggest over which range(s) to fit a power law. This approach, however, cannot be used when the number of data is small, and also, it does not give accurate estimates of the limits of the data interval. What we want is to characterize in a fully self-consistent way the distribution: this means to find exponent $\alpha$ and range $[x_m - x_M]$ with no a priori assumption. Here is the approach we followed.

As first step, we sort the $N$ data in increasing order. Then, we make a blind computation of the parameters $\alpha$, $x_m$ and $x_M$ using equation~(\ref{lS}) for the entire sample of the $N$ values, and the computed $\chi^2$ are stored. The parameters are left free but, a posteriori, we discard solutions with $\chi^2<1$.

Because we do not know the range over which the power-law distribution extends, this first solution found with all the $N$ values is not necessarily the most correct one. And it is also possible that no solution at all is found. So, we repeat the fit keeping all the data but the first one, again storing the $\chi^2$. The procedure is then continued removing one by one the smallest values, until we are left with only the last three data, with which a fit is not possible anymore. Thus, the fit is done over the entire range $[x_1 - x_N]$, then in the ranges $[x_2 - x_N]$, $[x_3 - x_N]$, and so on. At the end we have a collection, at most $N-4$ values, of $\chi^2$. We call this set of solutions $s_N$.

The best solution is expected to have the smallest $\chi^2$ but since each best-fit was derived with a different number of data, we have to compute the reduced $\chi^2$. This operation is not easy because for non-linear fits the number of degrees of freedom (\textit{dof}) is not known a priori. A common strategy is to set \textit{dof} equal to $N-n$, if $N$ is the number of points and $n$ is the number of parameters derived from the fit. But this is not always true for linear models, while for non-linear models \textit{dof} is essentially unknown \citep{andrae}.

A good example for understanding why \textit{dof} cannot always be derived assuming $N-n$ is given by the graybody emission $I_\nu$ \citep[e.g.,][]{gbody}
\begin{equation}
I_\nu=(1-\text{e}^{-\tau_\nu})B_\nu(T)\label{gbody}
\end{equation}
where $\tau_\nu$ is the optical depth and $B_\nu(T)$ is the Planck distribution at temperature $T$. In this case, the number, and even the physical meaning of the parameters depend on the different regimes $\tau\ll1$, $\tau\sim1$ or $\tau\gg1$. If we fit $I_\nu$ over a large range of frequencies, the number $n$ of parameters changes with $\nu$. For instance, in the region where $\tau\gg1$ we have $I_\nu\rightarrow B_\nu(T)$ and there is no longer a dependency on $\tau_\nu$.


Keeping in mind that for non-linear models computing $\chi_\text{red}^2$ is not possible \citep{andrae}, in the following examples we make the working hypothesis that the number of parameters is 3 and stays constant in all cases, so that, for each range $[x_k,x_N]$ with $1\le k\le N-4$, we tentatively compute $\chi_\text{red}^2=\chi^2/(k-3)$. Among the $N-4$ solutions we look for the one with the smallest $\chi_\text{red}^2$. In thiw way, we find the best solution for family $s_N$.

Because we do not know the range over which the power-law distribution extends, the first solution found with keeping the last value fixed, the most massive core for the CMF, as in Section~\ref{CMFP}, or the most energetic photon for the $\gamma$-ray spectrum, as in Section~\ref{Blazar}, is not necessarily the best one. So, we repeat the fit by removing the highest value among the observed data and using the remaining $N-1$ values. The family of solutions $s_{N-1}$ is built and the solution with the smallest $\chi_\text{red}^2$ is looked for, with now $1\le k\le (N-1)-4$, in the same way as done for the first case when we used all the $N$ values. The procedure is iterated until only the 4 smallest values are left whose family $s_4$ can have at most one solution.

At the end, for each family $s_i$ we have one or zero solutions. In fact, it is possible that a family does not have any solution: for instance, in the core mass function example, we do not expect a power-law distribution with $\alpha>1$ for the smallest masses, so it is not surprising that families $s_4$, $s_5$ up to $s_{144}$ (see next subsection) do not have any solution. But in the most favorable case we have $N-4$ solutions, one for each family $s_i$.


\subsection{The core mass function (CMF) in Perseus}\label{CMFP}
As an example of working with real data, in this section we use the sample of pre-stellar cores derived from Herschel observations in the Perseus star-forming region \citep{perseo}, to characterize the CMF with power-law distribution(s).

The CMF is the precursor of the Initial Mass Function (IMF) of the stars and can be described at small masses as a log-normal distribution, followed by a high-mass tail resembling a power law \citep{chabrier}. Even if the limit of validity of the log-normal at very small masses for the IMF is currently debated, see e.g., \citet{TPK} for $M\la0.1\,M_\odot$, such a functional form is often used to model the low-mass CMF. Further, CMFs are usually built with a few hundreds objects, up to $\sim500$ in Orion B \citep{orionB} and Orion A \citep[][Pezzuto et al., in preparation]{Takemura}, but often nearby molecular clouds contain a smaller number of cores. This makes it difficult or even impossible to distinguish if we are observing a pure log-normal or a power-law distribution \citep{swiftBeau}. Thus, with real data, we have to find $\alpha$ assuming
that the high-mass tail of the distribution follows a power law, without knowing where exactly the tail begins and ends.

\begin{figure}
	\includegraphics[width=\columnwidth]{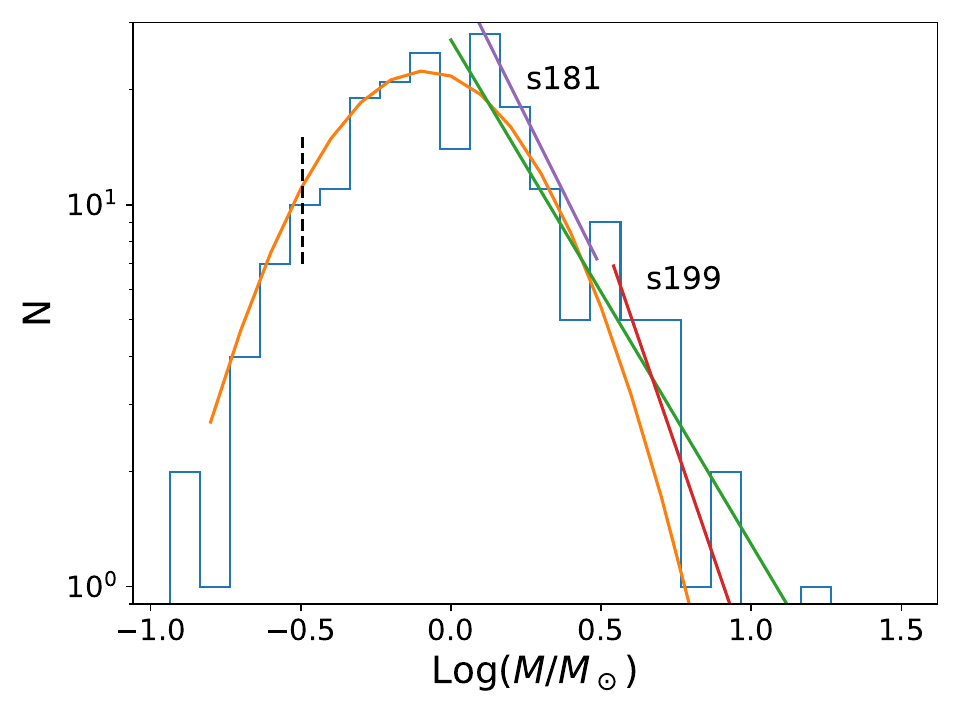}
    \caption{The core mass function ($\text{d}N/\text{d}\log M$) for 199 bound prestellar cores (cyan histogram) in Perseus. A bin size of $\log(M)=0.1$ was used. The orange line shows a log-normal fit, the green line the power-law fit, both derived by \citet{perseo} (see text for details). The dashed black line shows the completeness limit at $\sim0.32\,M_\odot$. Purple and red lines are the power-law distributions obtained with equation~(\ref{lS}) and discussed in the text \citep[adapted from][]{perseo}.}
    \label{CMF}
\end{figure}

In Perseus, \citet{perseo} found 199 starless gravitationally bound cores, that are alleged to be collapsing to form stars. For the log-normal part of the distribution the derived best-fit parameters are $\bar{m}=-0.086\pm0.028$, position of the peak, and $\sigma=0.347\pm0.028$, logarithmic dispersion. The high-mass tail was fit in the range $1\,M_\odot\le m\le16.742\,M_\odot$: the lower limit was chosen arbitrarily on the base of the histogram shape, while the higher limit was set equal to the highest core mass. The resulting exponent, found with the method by \citet{maizUbe}, was $\alpha=2.321\pm0.035$ (see Fig.~\ref{CMF}, green line).

To apply our technique, core masses are first sorted in increasing order and $\alpha$, $x_m$ and $x_M$ are computed for the entire sample of 199 pre-stellar cores. As done for simulations, $\alpha_\text{S}$ is computed to estimate $\alpha_\text{T}$ that, in turn, is used as first estimate of $\alpha_\text{L}$. First guesses of $x_m$ and $x_M$ are $0.9\mu_1$ and $1.1\mu_2$, with $\mu_1=\text{min}(x_i)$ and $\mu_2=\text{max}(x_i)$, respectively. The explored range for the parameters are: $-4\le\alpha\le+4$, $0.5\mu_1\le x_m\le2\mu_1$ and $0.5\mu_2\le x_M\le2\mu_2$.

We repeat the fit removing one by one the smallest masses, until we are left with only the last three most massive cores (family of solutions $s_{199}$). The solution with the smallest $\chi_\text{red}^2$ is stored and the procedure is repeated for family $s_{198}$, $s_{197}$ and so on.

The smallest $\chi_\text{red}^2$ is found for family $s_{181}$, meaning that the 18 most massive cores were not considered. Within this family, the best-fit is found after removing the 95 least massive cores leaving 77 cores in the range 1.02~$M_\odot$ -- 3.13~$M_\odot$. The parameters of the solution, shown in Fig.~\ref{CMF} as the purple line labelled $s_{181}$ are $\alpha=2.576\pm0.077$, $x_m=1.0583\pm0.0098\,M_\odot$ and $x_M=3.350\pm0.053\,M_\odot$ with $\chi_\text{red}^2\sim1$. This $\alpha$ is numerically close to Salpeter’s one, 2.35, differing from it, however, by at most $3\sigma$. The value $x_m\sim1.06\,M_\odot$ corresponds to a stellar mass of $\sim0.35\,M_\odot$, assuming a star formation efficiency per single core of 0.3 \citep{perseo}, which means that the final mass of stars is about one third of the core mass, close to the value 0.5~$M_\odot$ that marks the low-mass limit of the second power law in \citet{kroupa_jerabkova_2021}.
 
\begin{figure}
	\includegraphics[width=\columnwidth]{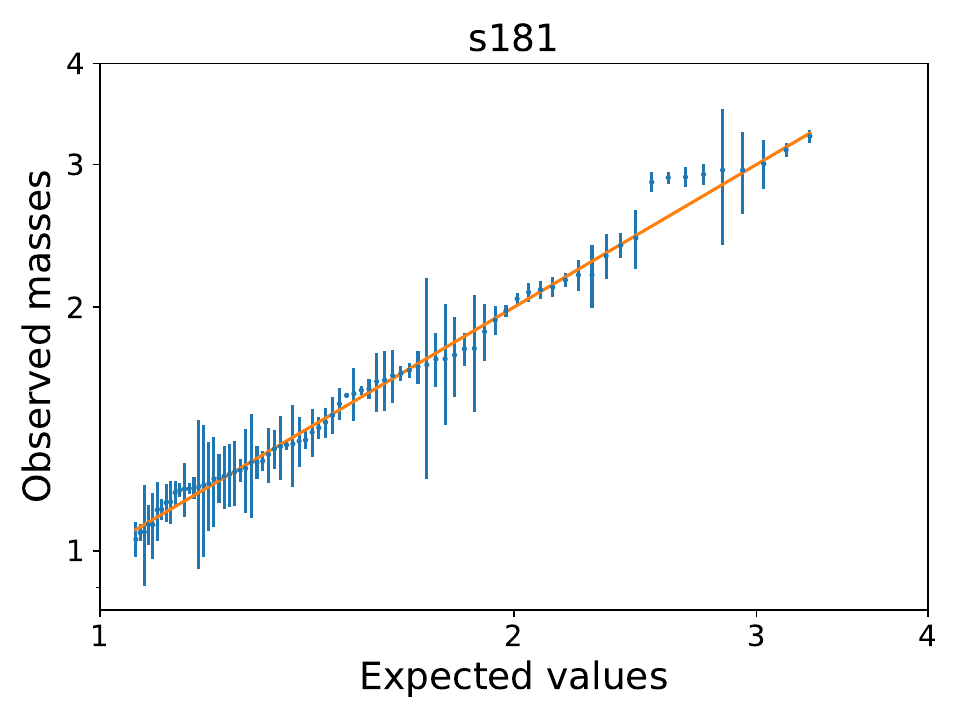}
	\includegraphics[width=\columnwidth]{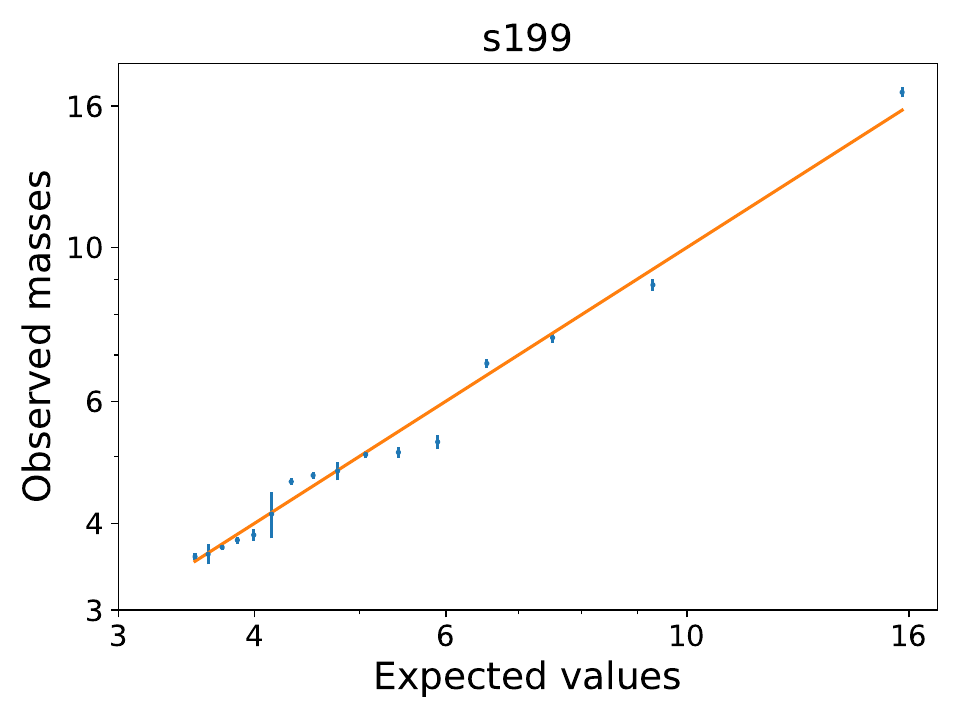}
    \caption{Observed masses of pre-stellar cores in Perseus vs. expected values $\bar\xi_i$ computed with equation~(\ref{expV}). Both axes in $M_\odot$. Top panel: solution $s_{181}$ where $\bar\xi_i$ are derived with $\alpha=2.58$, $x_m=1.06\,M_\odot$ and $x_M=3.35\,M_\odot$; the observed values are 77 cores with masses between 1.02~$M_\odot$ and 3.13~$M_\odot$. Bottom panel: solution $s_{199}$ with now $\alpha=3.39$, $x_m=3.48\,M_\odot$ and $x_M=33.4\,M_\odot$; the observed values are the 16 most massive cores. In both panels the orange line shows the ideal case of observed masses equal to expected values.}
    \label{expFit}
\end{figure}

Note that since our method works on unbinned data, a histogram plot is not the best way to show the solution. Thus, in the top panel of Fig.~\ref{expFit}, we compare the expected masses $\xi_i$, as derived from equation~(\ref{expV}), with the measured core masses, both sets in $M_\odot$. This way to compare expected values with observed masses is very similar to the PP (probability-probability) plot used in \cite{MK2009} as a test of the power-law hypothesis. The orange line in Fig.~\ref{expFit} shows the ideal case in which observed data are in perfect agreement with expected data. The main difference with respect to PP plot is that we do not use the cumulative function in the figure, so that the axes do not range from 0 to 1.

By looking at Fig.~\ref{CMF}, one may wonder if it is possible to make a fit using the remaining 18 most massive cores that were excluded to build family $s_{181}$. We thus made a second fit starting again by the most massive cores but now discarding solutions with $x_m<x_M(s_{181})=3.35\,M_\odot$ to avoid an overlap with the $s_{181}$ power law. There are nine possible solutions but three have negative $\alpha$, included the one having the smallest $\chi_\text{red}^2$, so we do not consider them. The remaing six solutions are summarized in Table~\ref{resCMF}.

\begin{table}
	\centering
	\caption{Parameters of the second power law.}
	\label{resCMF}
	\begin{tabular}{rrcccc}
$\chi_\text{red}^2$&N.&Range&$\alpha$&$x_m$&$x_M$\\
&&($M_\odot$)&&$(M_\odot)$&($M_\odot$)\\
9.22&16&3.58 - 16.7&$3.39\pm0.04$&$3.48\pm0.02$&$33.4\pm3.5$\\
8.24&15&3.58 - 8.83&$3.32\pm0.12$&$3.49\pm0.02$&$10.0\pm0.3$\\
8.97&14&3.58 - 7.41&$3.37\pm0.17$&$3.49\pm0.03$&$8.0\pm0.2$\\
10.70&12&3.61 - 6.81&$2.91\pm0.30$&$3.54\pm0.03$&$6.7\pm0.1$\\
3.84&8&4.61 - 4.76&$1.8\pm1.4$&$3.43\pm0.07$&$5.1\pm0.1$\\
5.03&6&3.70 - 4.69&$2.5\pm1.7$&$3.58\pm0.04$&$4.9\pm0.1$\\
\end{tabular}
\end{table}

Among the possible solutions, we show in Fig.~\ref{CMF} and in the bottom panel of Fig.~\ref{expFit}, $s_{199}$ that has the highest ratio $\alpha/\Delta\alpha$: parameters are $\alpha=3.389\pm0.044$, $x_m=3.481\pm0.020\,M_\odot$, $x_M=33.4\pm3.5\,M_\odot$. This slope is clearly steeper than, and incompatible with, Salpeter’s exponent of 2.35. It should be noted, however, that there are hints of possible steep $\alpha$ in the high-mass tail of the IMF. Assuming \citep{kroupa_jerabkova_2021} different slopes, $\alpha_1$, $\alpha_2$ and $\alpha_3$, for the ranges in mass $0.07\la M/M_\odot<0.5$, $0.5<M/M_\odot\le1$ and $M/M_\odot>1$, respectively, \citet{kroupa2001} proposed $\alpha_2=\alpha_3=2.35$. Equal exponents, $\alpha_2=\alpha_3$, are found also by \citet{sollima}, but in the range 2.41 -- 2.68, depending on the star formation history adopted to derive the IMF. Higher slopes, up to 3, are suggested also for IMF of open clusters \citep{ebrahimi}. Finally, \citet{mor} infer $\alpha_3$ between 2.9 and 3.7 (depending on the choice of the extinction map) for the IMF in the solar neighborhood. It is out of the scope of this paper to discuss the IMFs derived by different authors; what is important to us is that the slope $\sim3.5$ we found, may be acceptable. Moreover, the condition $m>x_m=3.481\,M_\odot$ implies for the final IMF in Perseus, again assuming a star formation efficiency per single core of 0.3, $m\ga1.2\,M_\odot$, close to the limit $m/M_\odot>1$ valid for $\alpha_3$.

We also note that no solutions were found for families $s_4$, using the 4 lightest cores, up to $s_{144}$. The total number of families for which no solution was found is 41 out of 195.

By looking at Fig.~\ref{CMF} it is clear that our solutions cannot be derived from the inspection of the histogram whose shape, by the way, depends on the bin size given the small number of cores. The unique solution found by \citet{perseo}, green line in Fig.~\ref{CMF}, seems enough to describe the high-mass tail of our distribution even if, from the figure, it is not clear where the single power law may actually start or end.

Our method, on the contrary, not only suggests that the high-mass CMF can be described with two power laws, as was already suggested by \citet{enoch} with observations at 1.1 mm; also, it gives the limits of the two power laws that are derived along with the exponents. In the other approaches \citep[see, e.g.,][for multiple power laws in PDF of column density maps]{marinkova} one has first to derive $\alpha$ for different sets of $x_m$ and $x_M$ (but note that, often, the possibility to have $x_M$ finite is not considered), and then to apply a posteriori checks to select the best choice for the three parameters.

We conclude this section noting that modelling the CMF with a power-law probability distribution function as done in equation~(\ref{nx}), does not imply that the mass of the cores is the result of a stochastic process\footnote{Note, however, that \citet{1994A&A...287..517R} derived the Salpeter's index by assuming that the star formation is a completely random process}. In fact, \citet{adamsFat} showed that if the final mass of a star is physically determined by $n$ quantities distributed according to a power law, the resulting IMF will also follow a power law in the high-mass tail. Further, self-regulated star formation can be described in terms of power-law distributions \citep[e.g.,][]{YAN}, and a self-regulated process is clearly not random.


\subsection{The slope of the high-energy spectrum in the BL Lac source J0011.4+0057}\label{Blazar}
In high-energy astrophysics, when dealing with X- or $\gamma$-rays observations, the number of collected photons is generally small so that the energy of each single photon is measured. It is, thus, possible to derive the slope of a spectrum with the method presented in this paper. To this aim, we used datat of the blazar, J0011.4+0057, observed with \textsl{Fermi}-LAT instrument \citep{FERMILAT}. This source appears in the paper accompanying the \textsl{Fermi}-LAT fourth data release \citep{FSC4}, where it is classified as a flat-spectrum radio quasar.

We have downloaded from the archive\footnote{\url{https://fermi.gsfc.nasa.gov/cgi-bin/ssc/LAT/LATDataQuery.cgi}} data collected in the period starting on 13/3/2022 at 3:5:18 and ending on 9/9/2022 at 00:14:04. They consist of 85,185 photons with energies $E$ in the range $100\,\text{Mev}\le E\le97.8\,\text{Gev}$.

Following the instructions available in the archive, we select only photons with associated \texttt{Event Class} set to 128, recommended in the archive for most analyses with good sensitivity of point sources and moderately extended sources; and \texttt{Conversion Type} equal to 1 or 2, meaning that photons were detected in the Front- or Back-section of the Tracker, respectively. After this step, the number of photons reduces to 45,033.

Next, low-energy photons are discarded to minimize the presence of other sources within the point spread function (PSF) of the instrument. The PSF depends on photon energy as $E^{-0.8}$ with a 68\% containment radius $\sim5\degr$ at 100~Mev \citep{FERMILAT}. At an angular distance of $1\fdg73$ from our source, there is another blazar, J0016.2–0016, brighter than J0011.4+0057 up to 1~Gev. We have computed the energy at which the 99.9\% containment radius is smaller than $1\fdg73$. Assuming that the PSF can be approximated with a Gaussian, so that at 100~Mev the 99.9\% containment radius is $15\degr$, all the photons with energy $E<1484.21$~Mev are removed. In the end, we are left with 3,417 photons. The uncertainty on the energy is set to 2\% \citep{ackermann}.

\begin{figure}
	\includegraphics[width=\columnwidth]{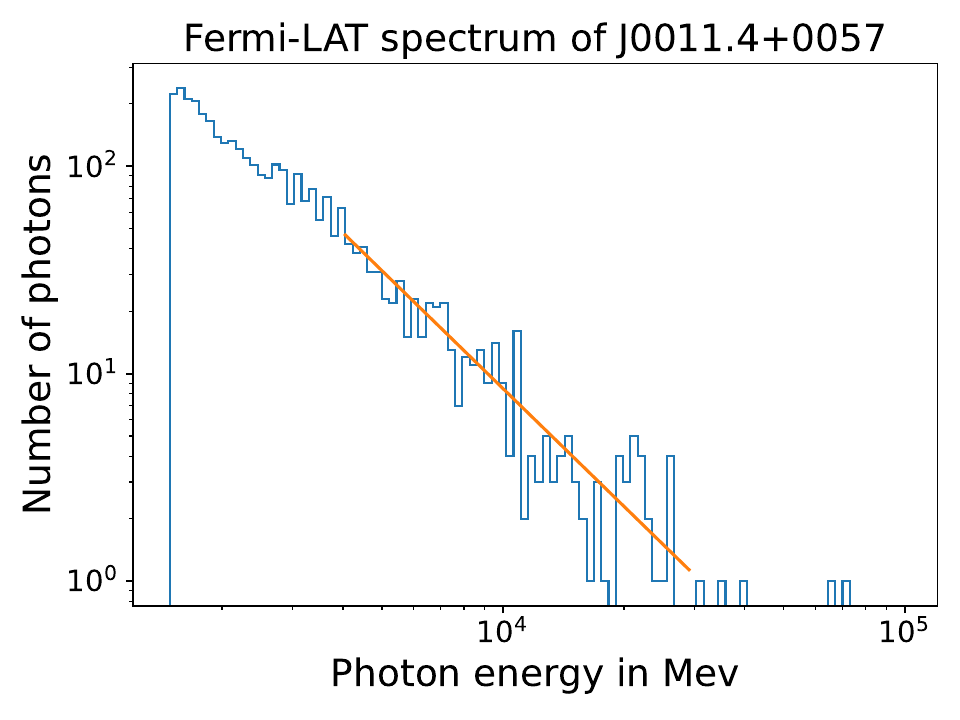}
    \caption{Blue histogram: \textsl{Fermi}-LAT $\gamma$-ray spectrum of the blazar J0011.4+0057 in the range $1484.21\,\text{Mev}\le E_\text{ph}\le97.8$~Gev. Photons with $E_\text{ph}<1484.21$~MeV were removed to avoid contamination from a close bright source. Orange line: our power-law solution with $\alpha=2.89$, $E_m=4.16$~Gev and $E_M=28.7$~Gev.}
    \label{gammaFit}
\end{figure}

We follow the same procedure adopted in the previous section to derive the slope of the CMF in Perseus. For each high-energy photon we look for the best-fit power law after removing one by one all the low-energy photons. The smallest $\chi_\text{red}^2$, of the order of 1, is found for the solution $s_{3410}$ using 538 photons in the range 4.15 - 26.3~GeV with parameters $\alpha=2.8871\pm0.0081$, $E_m=4,163.4\pm6.4$~MeV and $E_M=28.68\pm0.23$~GeV. The solution is shown in Fig.~\ref{gammaFit}.

\citet{FSC4} fit the observed $gamma$--ray spectra of the entire sample of sources in the \textsl{Fermi}-LAT archive with three different functional forms; best-fit parameters are also reported in the archive. One of the applied fits is a power law
\begin{equation}
\frac{\text{d}N}{\text{d}E}=K\left(\frac{E}{E_0}\right)^{-\Gamma}.
\end{equation}
For J0011.4+0057 the parameters are $K=(4.05\pm0.40)\times10^{-13}$~ph~cm$^{-2}$~Mev$^{-1}$~s$^{-1}$, $E_0=1,208$~Mev and $\Gamma=2.319\pm0.085$. A direct comparison of our $\alpha$ with $\Gamma$ is not easy because we selected photons with energy $E>1,484.21\,\text{Mev}>E_0$, so that the two fits were done over two different portion of the spectrum.

The 3,417 photons used to build the histogram in Fig.~\ref{gammaFit} were collected in almost 6 months. However, it is known that blazar can show variability on much shorter time scales, down to 3 – 4 days \citep{abdo}. For this reason, and because our technique can work on a much smaller number of photons, we split the entire observation in 16 bunches of 200 photons each, with the last 217 photons in the seventeenth block. The analysis is then performed for each subset, keeping the best-fit solution for each time interval.

\begin{figure}
	\includegraphics[width=\columnwidth]{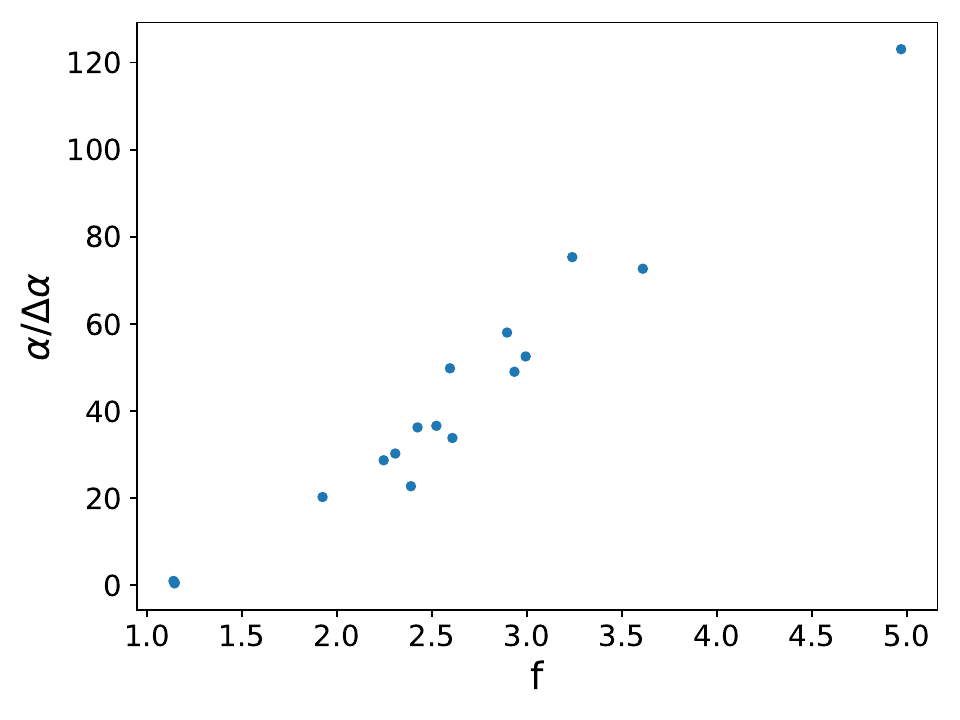}
	\includegraphics[width=\columnwidth]{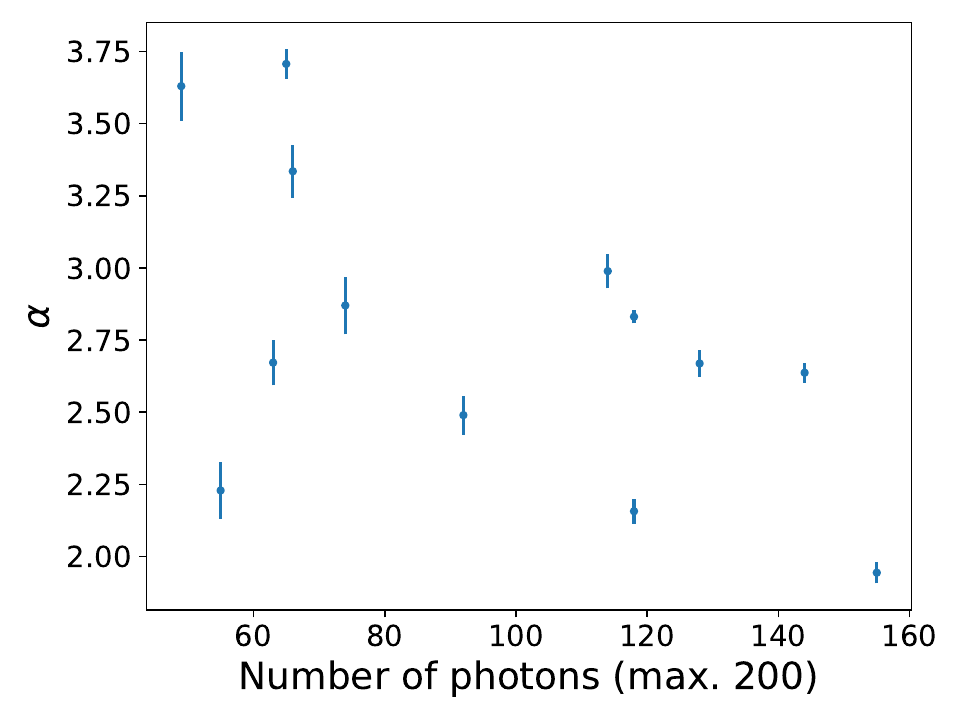}
    \caption{Top panel: signal to noise ratio $\alpha/\Delta\alpha$ for the slope of the 17 $\gamma$-rays spectra vs $f=E_M/E_m$ (blue points). Bottom panel: slopes, after removing the 3 values corresponding to $f<2$, vs. the number $N$ of used photons ($4\le N\le200$).}
    \label{sceltaAlpha}
\end{figure}

In the top panel of Fig.~\ref{sceltaAlpha} we show with blue points the 17 ratios $\alpha/\Delta\alpha$ vs. ratio $f=E_M/E_m$. By looking at the figure we tentatively assume that for $f<2$ the slopes are not reliable: three of them are indeed ill-measured ($\alpha=3.6\pm6.3$ for first block; $\alpha=1.0\pm1.1$ for sixth block; $\alpha=2.7\pm6.2$ for block 13). The fourth value has a high signal to noise ratio ($\alpha=4.66\pm0.23$ or $\alpha/\Delta\alpha\sim20$, fifth block) but it looks too steep \citep[slopes larger than 3.2 are not reported by][]{FSC4}. For comparison, we report here that both solutions found in the previous section for the Perseus CMF have $f>3$ and $\alpha/\Delta\alpha>30$.

\begin{figure*}
\begin{tabular}{ccc}
	\includegraphics[scale=.33]{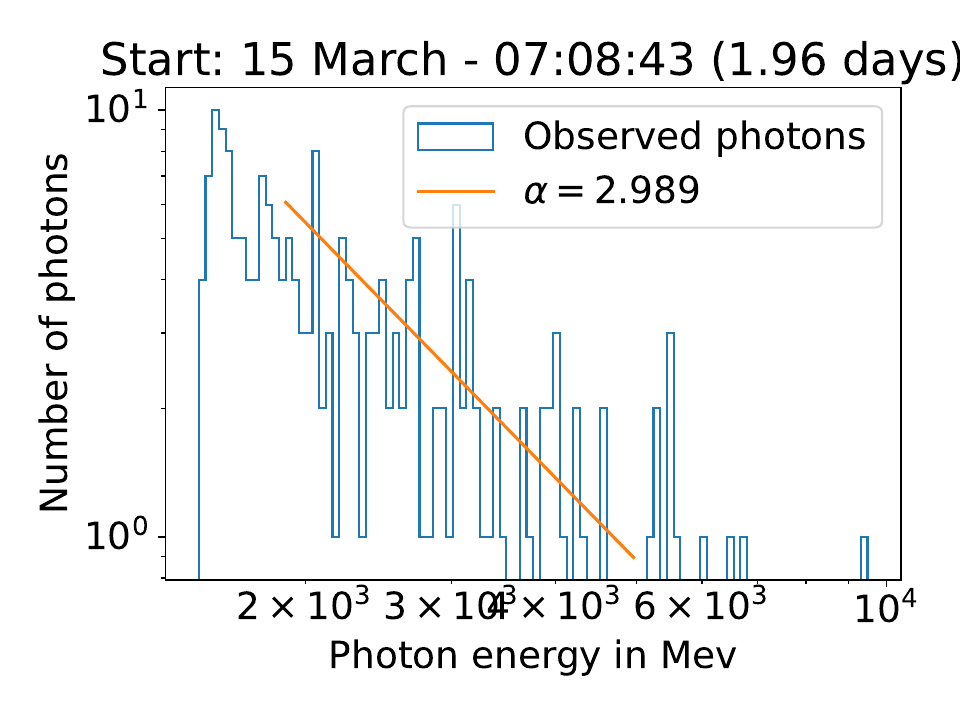}&
	\includegraphics[scale=.33]{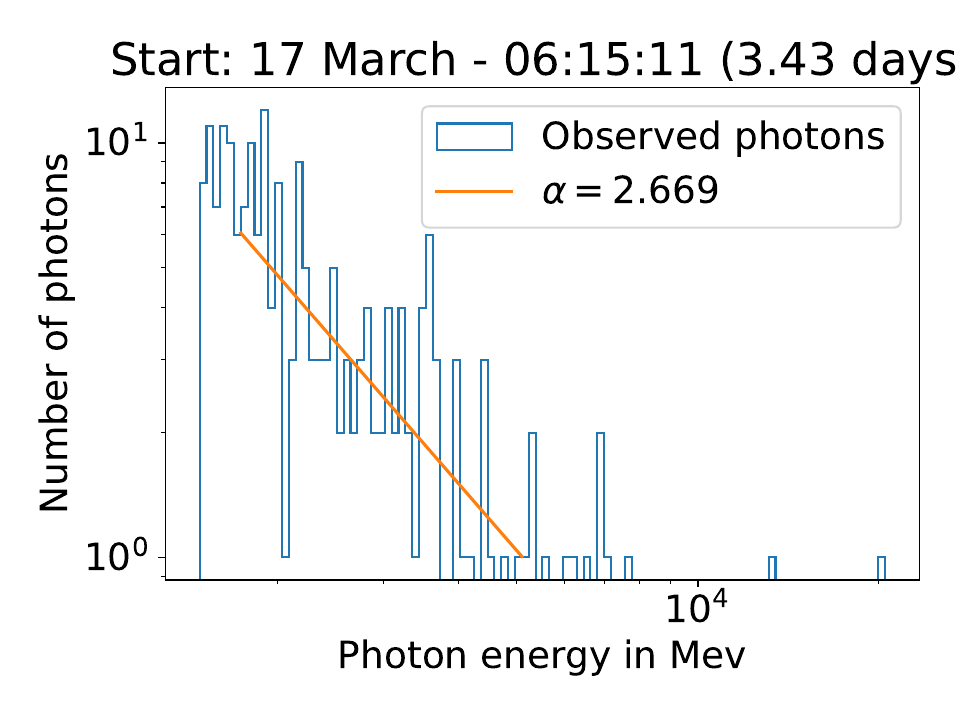}&
	\includegraphics[scale=.33]{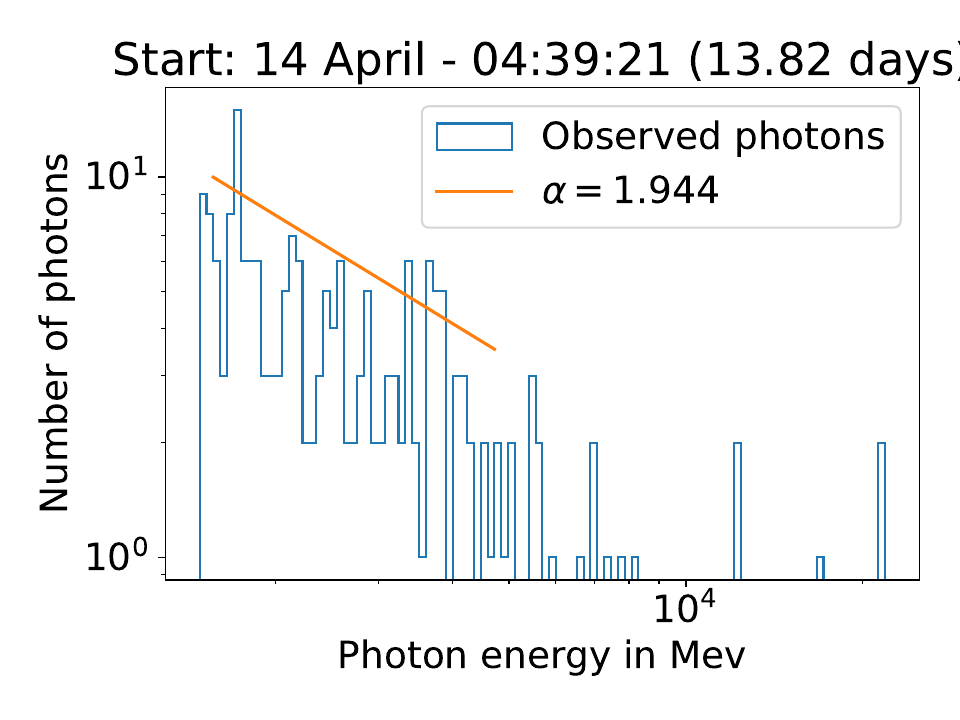}\\
	\includegraphics[scale=.33]{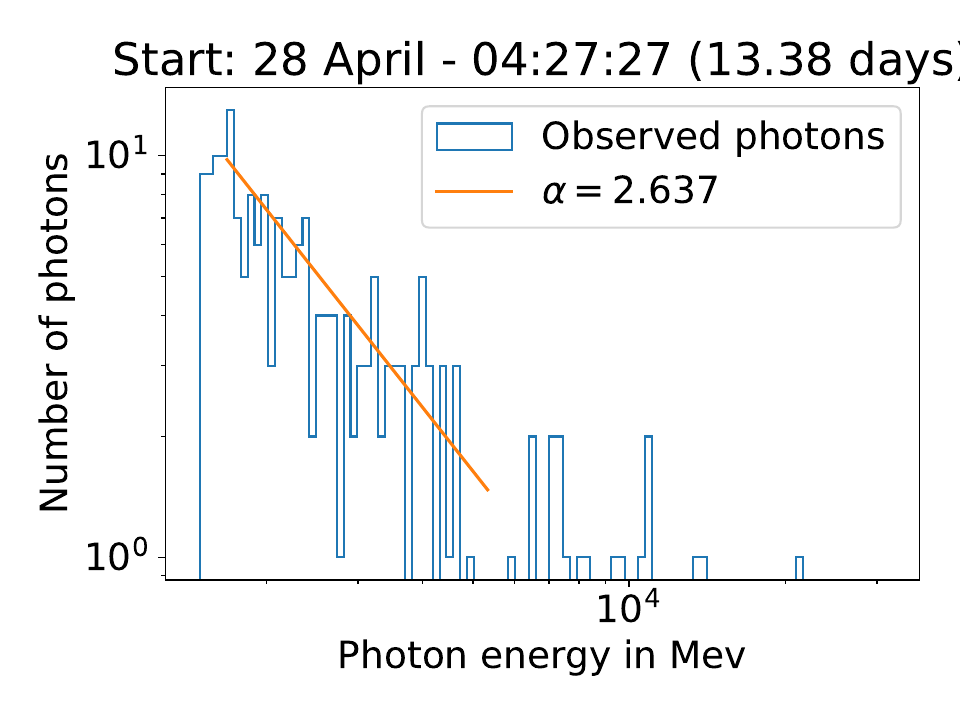}&
	\includegraphics[scale=.33]{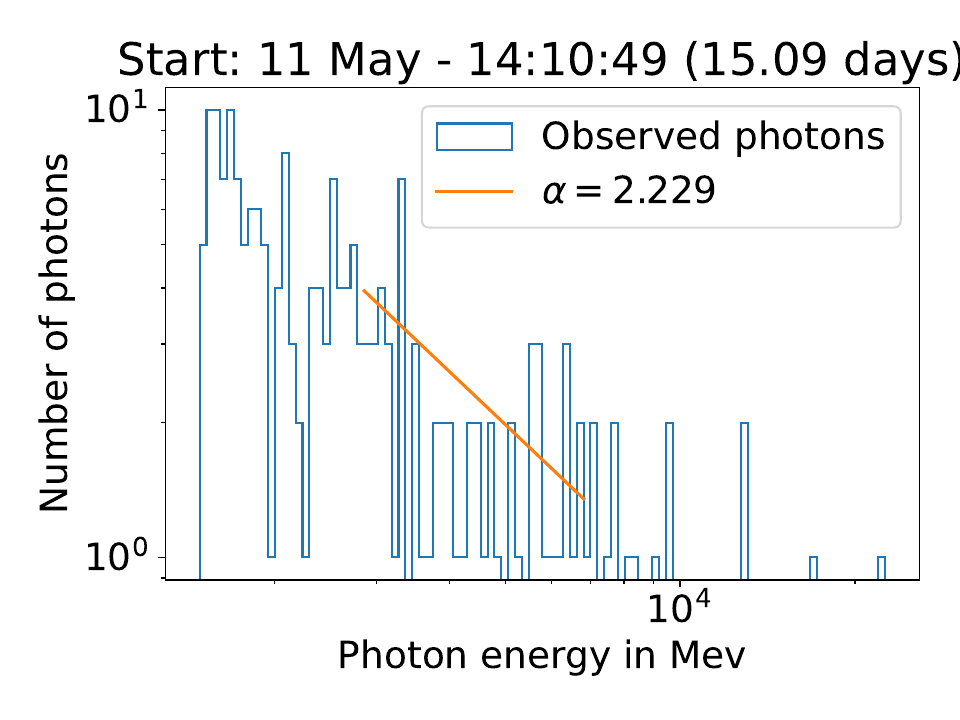}&
	\includegraphics[scale=.33]{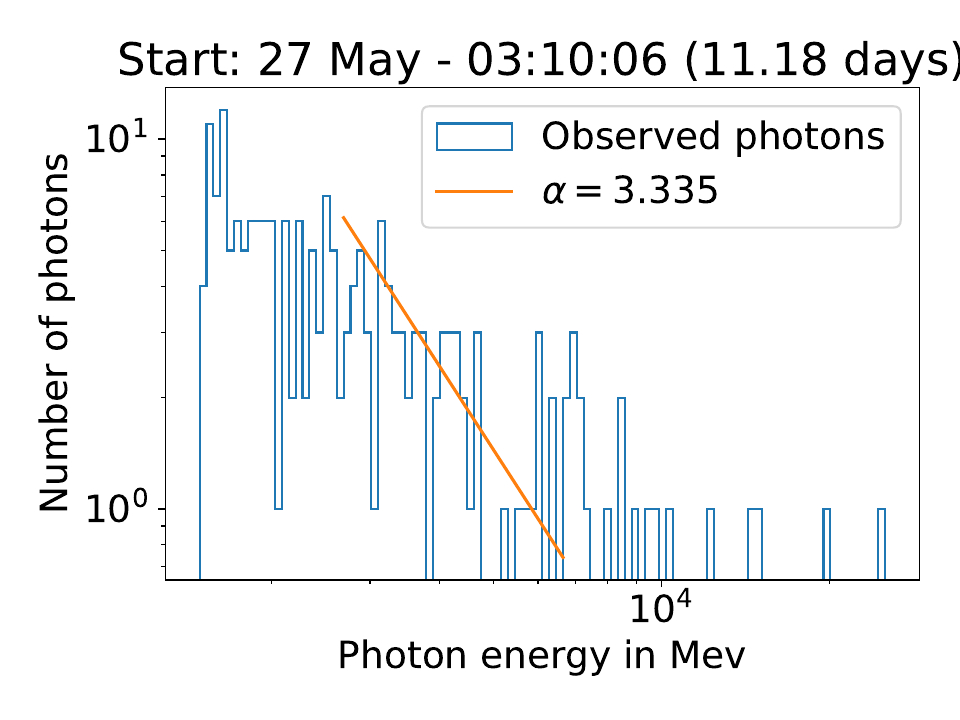}\\
	\includegraphics[scale=.33]{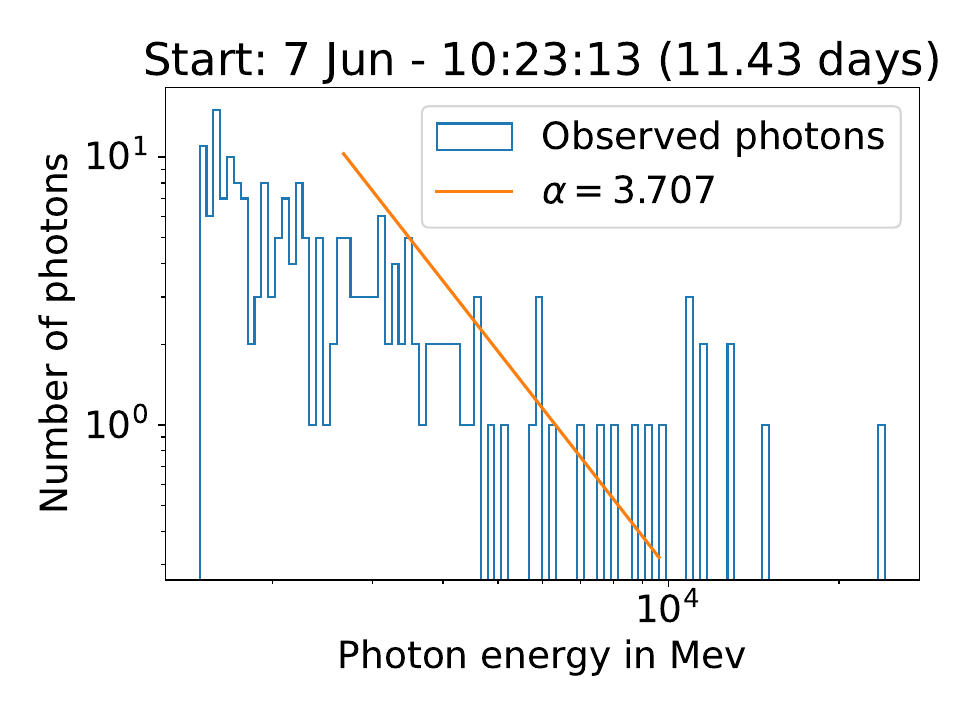}&
	\includegraphics[scale=.33]{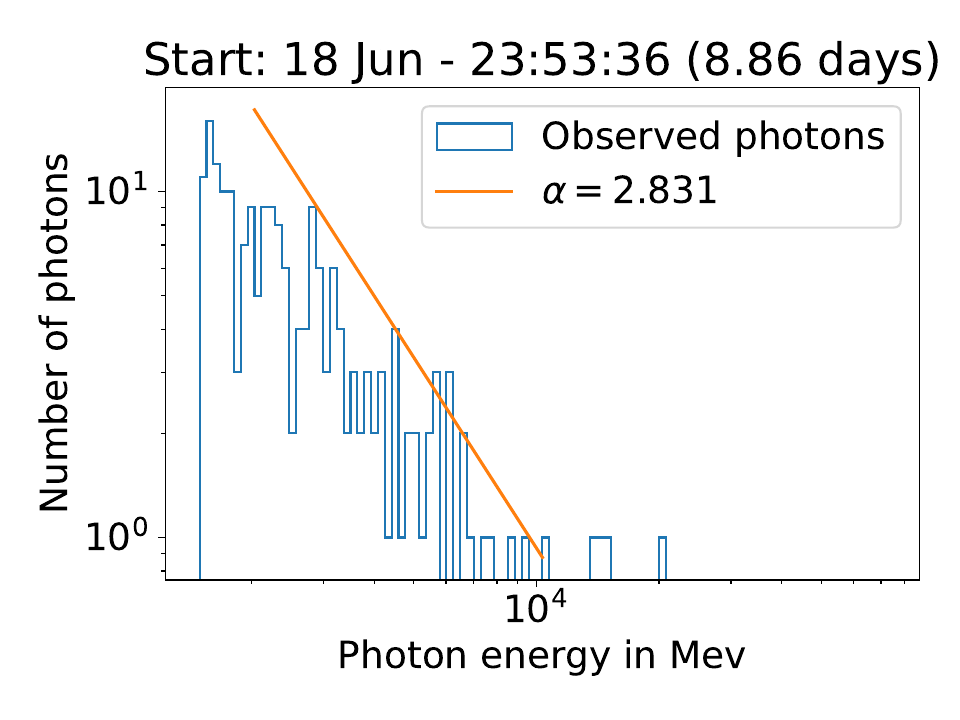}&
	\includegraphics[scale=.33]{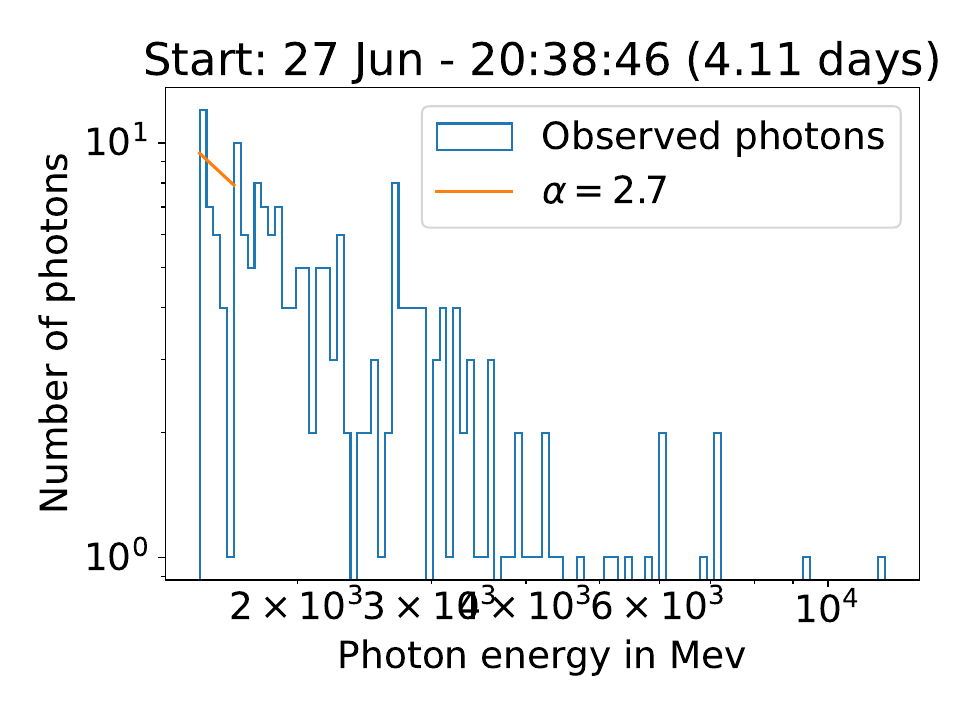}\\
	\includegraphics[scale=.33]{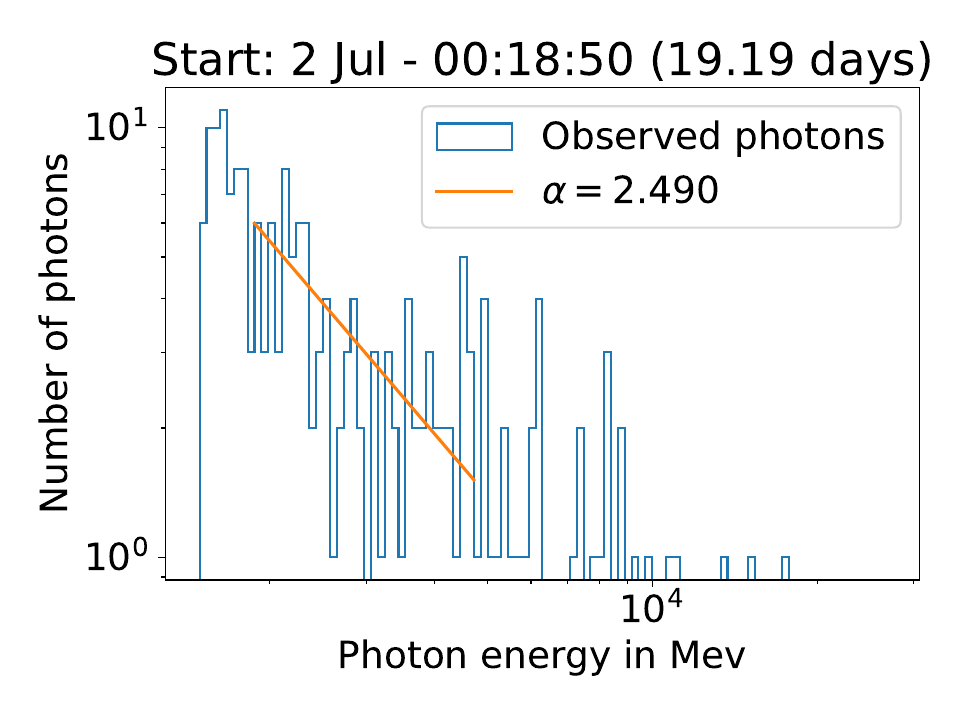}&
	\includegraphics[scale=.33]{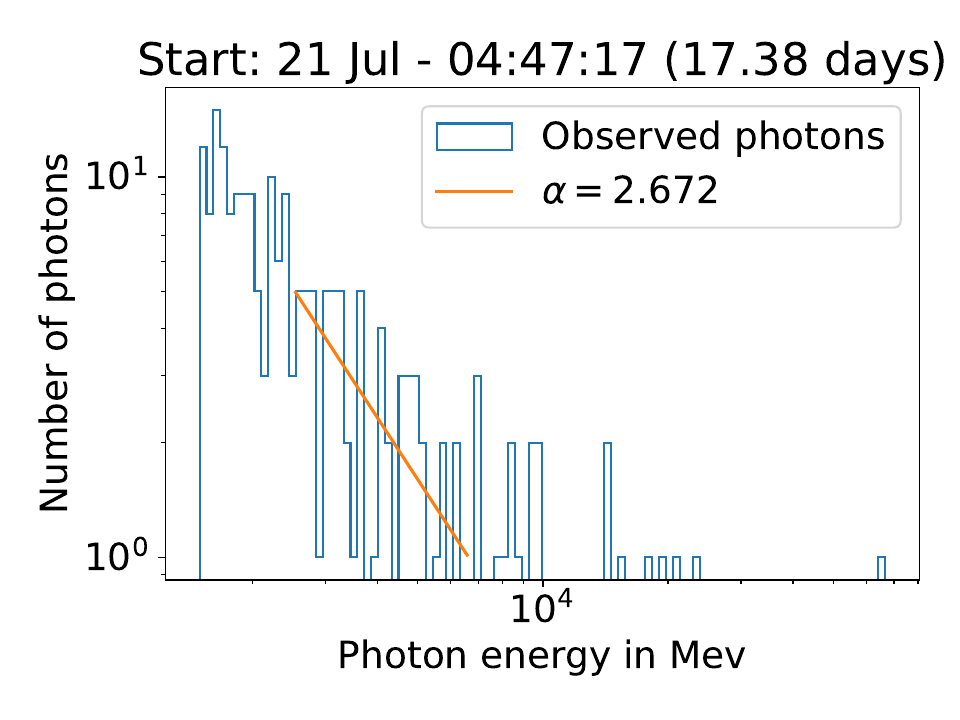}&
	\includegraphics[scale=.33]{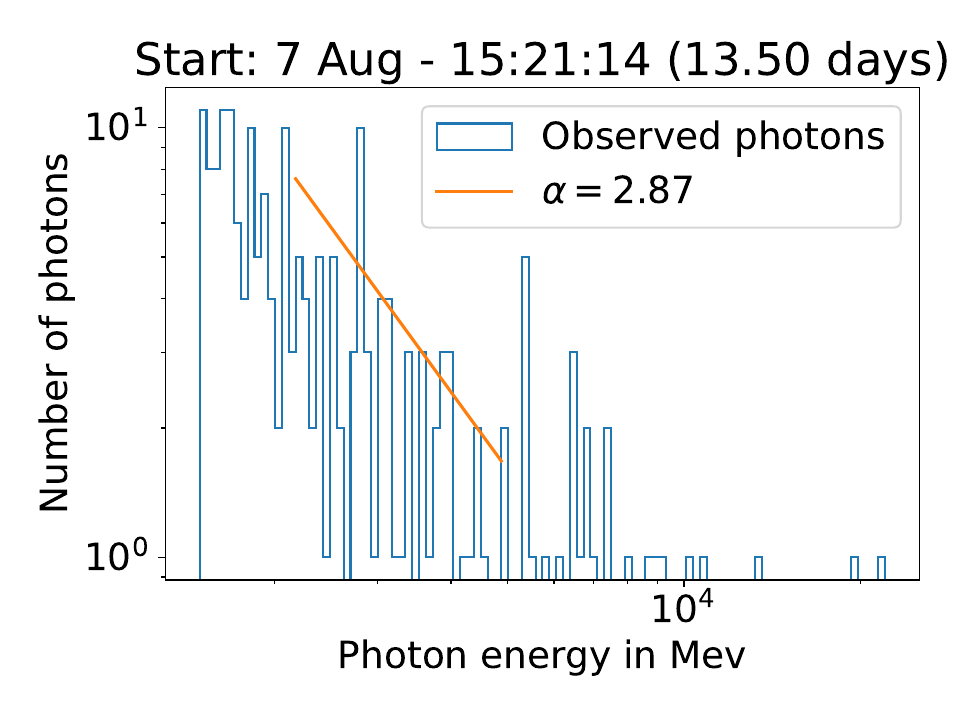}\\
	\includegraphics[scale=.33]{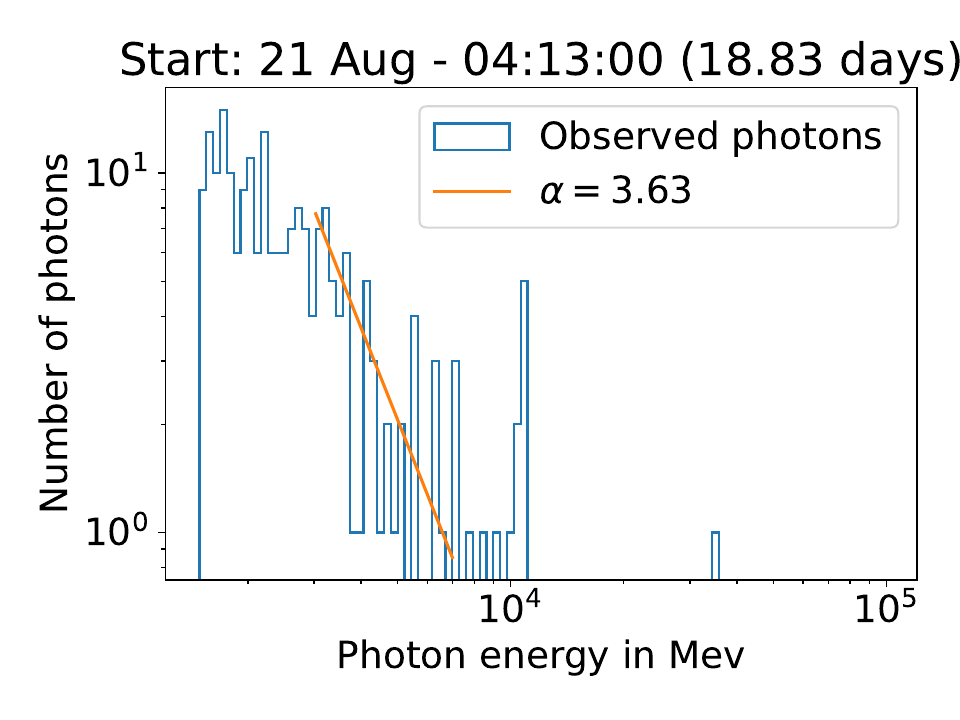}\\
\end{tabular}
    \caption{The 13 time-resolved spectra, see the bottom panel of Fig.~\ref{sceltaAlpha}, of the $\gamma$-ray source J0011.4+0057, obtained with 200 photons each (217 for the last block). Start time is the arrival time of the first photon of the spectrum; in parenthesis we give the time interval necessary to collect the 200 photons.}
    \label{spettriGamma}
\end{figure*}

In the bottom panel of Fig.~\ref{sceltaAlpha} we show the remaining 13 values of $\alpha$ vs. $N$, the number of used photons. There is a faint indication of decreasing $\alpha$ with increasing $N$ but this result may arise from the arbitrariness in splitting the entire observation in 200-photons blocks: some blocks could mix different physical states of the source. We present this analysis only to demonstrate that the spectrum can be fit also with few photons. The physical analysis of the results, however, is out of the scope of this paper. The 13 spectra are shown in Fig.~\ref{spettriGamma} along with the best-fit solutions.

%
%

\section{Conclusions}\label{conclusioni}
In this paper, we presented a new tool to derive the exponent $\alpha$ of a power-law distribution, $x^{-\alpha}$, and the limits of the interval over which the distribution can be fit. We considered distributions defined both over closed intervals $[x_m,x_M]$ (truncated distribution), and over open intervals. In the latter case, if $\alpha>1$ the interval is $[x_m,\infty)$, while for $\alpha<0$ the range of validity is $(0,x_M]$. For $0\le\alpha\le1$ only truncated distributions can be normalized.

Our method is based on the hypothesis that, more frequently, the set of $N$ observed values $\{x_i\}$ obeying a power-law distribution will populate the interval (open or closed) with some regularity. Under this assumption, the entire interval can be split in $N$ sub-intervals of equal probability to contain one $x_i$. For each sub-interval the expectation value $\xi_i$ is then computed. The set of $N$ expectation values depends on the three parameters of the distribution; by minimizing the differences between these $\xi_i$ and the observed values $x_i$, through a (non-linear) least-squares fitting algorithm, we can derive the unknown parameters that best describe the set of $\{x_i\}$.

We compared our method with two already known techniques: the maximum likelihood estimator (MLE) derived for open intervals (that we named standard MLE or SMLE), and for closed intervals (truncated MLE or TMLE). Starting from a set of 1,000 series of 10,000 uniform random numbers, first we built power-law simulated distributions over the open interval $[0.8,\infty)$ with $\alpha=\{1.5, 2.0, 2.5, 3.5\}$. The reason for choosing these values is explained in the text, here we note that for these kind of distributions $x_m$ acts as a scale factor, so that its precise value is not important. Then we derived $\alpha$ using the two MLE estimators ($\alpha_\text{S}$ and $\alpha_\text{T}$ for SMLE and TMLE, respectively) and the three parameters $\alpha$ ($\alpha_\text{L}$), $x_m$ and $x_M$ using our own technique. We found that:
\begin{enumerate}
\item for $\alpha=3.5$, $\alpha_\text{S}$ and $\alpha_\text{T}$ require $\sim30$ values $x_i$ to determine the slope $\alpha$ with a precision of 10\% (10\% criterion in the following), or better. When $\alpha$ decreases, $N$ decreases as well: for $\alpha=1.5$, 10 values are enough to meet the 10\% criterion;
\item the main limitation on deriving a correct $\alpha_\text{S}$ is the lack of knowledge of $x_m$, so that for small $N$ $\alpha_\text{S}>\alpha$. We give a mathematical interpretation of this inequality. On the contrary, $\alpha_\text{T}<\alpha$ for small $N$;
\item our method performs as well as the other two, even if the 10\% criterion is met with a slightly higher $N$, up to 50 for $\alpha=3.5$. Note, however, that the convergence of the mean to the true value is faster with our methods than with the other two, the higher $N$ depends on the larger scatter around the mean. This behaviour is found also for $x_m$: the mean converges rapidly to the true value, but with a large standard deviation. Opposite to $\alpha_\text{L}$, $x_m$ requires higher $N$, up to $\sim140$, for smaller $\alpha$ while only 5 values are necessary for $\alpha=3.5$.
\end{enumerate}

As we write in the text, simulations extending over open intervals, especially for small $\alpha$, are not very realistic. More interesting is the case of truncated power-law distributions, where the variable $x$ is bounded on both sides: $x_m\le x<x_M$. We started from the same set of uniform random numbers simulations as in the previous case, deriving now power-law distributions over the interval $[0.8,40)$. As before, $x_m$ acts as a scale factor, while $x_M$ enters only through the ratio $f=x_M/x_m$. With our choice of parameters we have $f=50$. Our results are:
\begin{enumerate}
\setcounter{enumi}{3}
\item as $\alpha$ decreases, $\alpha_\text{S}$ becomes a worse estimate of $\alpha$. For $\alpha=1.5$, $\alpha_\text{S}$ never converges to $\alpha$, attaining the limiting value of 1.74. What we found, but exploring a limited parameter space, is that $\alpha_\text{S}$ stops being a good estimate of $\alpha$ when $f^{1-\alpha}\ga0.02$;
\item $\alpha_\text{T}$ works well also for truncated distributions, requiring less than 40 values to reach the 10\% convergence. The drawback of this MLE is that now two estimates are required: one for $x_m$ and one for $x_M$;
\item $\alpha_\text{L}$ performs as well as $\alpha_\text{T}$, requiring just slightly higher $N$, about 60 in the worst case, to converge to $\alpha$;
\item for $x_m$ the estimates follow the same trend with $\alpha$ as in the open-interval case, but requiring much less data: only 50 when $\alpha=1.5$;
\item for $x_M$, the convergence strongly depends on $\alpha$ with only 8 values necessary to estimate the true $x_M$ when $\alpha=1.5$. But for $\alpha=3.5$ we need $\sim5,000$ data to meet the 10\% criterion. We demonstrate that this bad performance is caused by the steepness of the distribution: for large $\alpha$, long series of data are necessary to have high values of $x$ so that our estimates of $x_M$ are, in fact, in line with the simulated data.
\end{enumerate}

Finally, we applied our method to real data:
\begin{enumerate}
\setcounter{enumi}{8}
\item for the slope of the core mass function (CMF) in Perseus we find as best solution $\alpha=2.576\pm0.077$, $x_m=1.0583\pm0.0098\,M_\odot$ and $x_M=3.350\pm0.053\,M_\odot$. This slope is a bit steeper than Salpeter’s exponent of 2.35. The set of data also suggests a possible second solution with $\alpha=3.389\pm0.044$, $x_m=3.481\pm0.020\,M_\odot$ and $x_M=33.4\pm3.5\,M_\odot$. This high value of $\alpha$ is compatible with the finding that the distributions of the star high-mass tail shows slopes as high as 3.7, when the stars initial mass function (IMF) is parametrized with three power laws. The statistical significance of this second solution is, however, small;
\item for the spectrum of the $\gamma$-ray source J0011.4+0057, a blazar of type flat-spectrum radio quasar observed with the \textsl{Fermi} satellite, we compute a slope $2.8871\pm0.0081$, with limits $E_m=4,163.4\pm6.4$~MeV and $E_M=28.68\pm0.23$~GeV. Slopes are found for time resolved spectra too, after splitting the entire collection of 3,417 most energetics photons in smaller sets, 200 photons each, sorted in temporal order. The derived slopes are between $\sim2$ and $\sim3.75$.
\end{enumerate}

We conclude that our technique is a promising tool to describe power-law distributions giving at the same time the three unknowns $\alpha$, $x_m$ and $x_M$.

\section*{Acknowledgements}
The authors thank the anonymous referee for providing important additional references on the subject of power-law distributions. The authors acknowledge support from the European Research Council Synergy Grant ECOGAL (Grant: 855130). RSK also acknowledges funding from the Deutsche Forschungsgemeinschaft (DFG) via the Collaborative Research Center ``The Milky Way System'' (SFB 881 – funding ID 138713538 – subprojects A1, B1, B2 and B8), from the Heidelberg Cluster of Excellence (EXC 2181 - 390900948) “STRUCTURES”, funded by the German Excellence Strategy, and from German Ministry for Economic Affairs and Climate Action in project “MAINN” (funding ID 50OO2206). We wish to thank IAPS collegues Dr. Giovanni Piano for valuable hints on accessing and retrieving data from the \textsl{Fermi}-LAT archive; Dr. Manuela Magliocchetti for suggesting example of power laws in cosmological context; and Dr. Fabrizio De Angelis for technical support. Appendix~\ref{derAlpha} benefited from website \url{http://www.wolframalpha.com/} by Wolfram Alpha LLC. 2009. Wolfram$|$Alpha. 

\section*{Data Availability}

All the python scripts used in the paper to compute $\alpha$, as well as the simulated data, are available on-line at \url{ https://drive.google.com/drive/folders/0ANHuArSAlRAJUk9PVA}.



\bibliographystyle{mnras/mnras}
\bibliography{msRev1} 




\appendix

\section{Additional formulae for the least-squares method}\label{addForm}
\subsection{Expected values for particular exponents}\label{specCas}
In this appendix we give the set of equations~(\ref{defXi}) and (\ref{expV}) for $\alpha=1$ and $\alpha=2$. There are at least two reasons to do that: first, the problem under study may suggest that the solution is $\alpha\sim1$ or 2, so that these values can be used as initial estimates; second, it is not impossible, even if quite unlikely, that during the fitting procedure the entire fractional part of $\alpha$ becomes zero and $\xi_i$ are to be computed for $\alpha\sim1$ or 2.

Let us start with $\alpha=2$: in this case equations~(\ref{defXi}) are still valid and become
\begin{equation}
\xi_i=x_m\left[1-\frac{i}N\left(1-f^{-1}\right)\right]^{-1}\,\,\,1\le i\le N-1
\end{equation}
while equation~(\ref{pxt}) is
\begin{equation}
p(x)\text{d}x=\frac{x_M}{f-1}x^{-2}\text{d}x.
\end{equation}

The $\bar\xi_i$ are now
\begin{equation}
N\int_{\xi_{i-1}}^{\xi_i}xp(x)\text{d}x=N\int_{\xi_{i-1}}^{\xi_i}\frac{x_M}{f-1}x^{-1}\text{d}x=\frac{Nx_M}{f-1}\ln\frac{\xi_i}{\xi_{i-1}}
\end{equation}
with $1\le i\le N$, $\xi_0=x_m$, $\xi_N=x_M$; finally
\begin{equation}
\bar\xi_i=\frac{Nx_M}{f-1}\ln\frac{1-\frac{i-1}N(1-f^{-1})}{1-\frac{i}N(1-f^{-1})}\,\,\,1\le i\le N.
\end{equation}
or, equivalently,
\begin{equation}
\bar\xi_i=\frac{Nx_M}{f-1}\ln\left[1+\frac{f-1}{Nf-i(f-1)}\right].
\end{equation}

If $\alpha=1$ we start from equation~(\ref{pxt1}) so that
\begin{equation}
\xi_i=x_m\left[\text{e}^\frac{\ln f}N\right]^i
\end{equation}
from which
\begin{equation}
\bar\xi_i=\frac{Nx_m}{\ln f}\left[\text{e}^\frac{\ln f}N-1\right]\left(\text{e}^\frac{\ln f}N\right)^{i-1}.
\end{equation}

\subsection{The derivatives of $\bar\xi_i$}\label{derAlpha}
Because equations~(\ref{lS}) form a non-linear system, the solution cannot be found analytically. In general, the algorithms that does the numerical job require the derivatives $\partial\bar\xi_i/\partial p_j$. For the reader’s convenience, we report the derivatives here.

We start by writing
\begin{equation}
\bar\xi_i=q_1q_2q_3
\end{equation}
with
\begin{align}\nonumber
q_1&=\frac{\alpha-1}{2-\alpha},\\
q_2&=\frac{Nx_m}F,\\\nonumber
q_3&=t_i^\frac{2-\alpha}{1-\alpha}-t_{i-1}^\frac{2-\alpha}{1-\alpha}
\end{align}
where $F=1-f^{1-\alpha}$, $t_i=1-iF/N$ and $t_{i-1}=1-(i-1)F/N$. Thus
\begin{align}\nonumber
\frac{\partial\bar\xi_i}{\partial x_m}&=\frac{\bar\xi_i}{q_2}\frac{\partial q_2}{\partial x_m}+\frac{\bar\xi_i}{q_3}\frac{\partial q_3}{\partial x_m},\\
\frac{\partial\bar\xi_i}{\partial x_M}&=\frac{\bar\xi_i}{q_2}\frac{\partial q_2}{\partial x_M}+\frac{\bar\xi_i}{q_3}\frac{\partial q_3}{\partial x_M},\\\nonumber
\frac{\partial\bar\xi_i}{\partial\alpha}&=\frac{\bar\xi_i}{q_1}\frac{\partial q_1}{\partial\alpha}+\frac{\bar\xi_i}{q_2}\frac{\partial q_2}{\partial\alpha}+\frac{\bar\xi_i}{q_3}\frac{\partial q_3}{\partial\alpha}.
\end{align}

Computing the first two derivatives is long but not difficult
\begin{align}
\frac{\partial\bar\xi_i}{\partial x_m}=&\frac{N[1+(\alpha-2)f^{1-\alpha}]}{F^2}\frac{\bar\xi_i}{q_2}+\\
&+\frac{\alpha-2}{Nx_m}f^{1-\alpha}\left[i\left(t_i\right)^\frac1{1-\alpha}-(i-1)\left(t_{i-1}\right)^\frac1{1-\alpha}\right]\frac{\bar\xi_i}{q_3}\nonumber
\end{align}
and
\begin{equation}
\frac{\partial\bar\xi_i}{\partial x_M}=\frac{N(1-\alpha)}{f^\alpha F^2}\frac{\bar\xi_i}{q_2}+\frac{2-\alpha}{Nx_m}f^{-\alpha}\left[it_i^\frac1{1-\alpha}-(i-1)t_{i-1}^\frac1{1-\alpha}\right]\frac{\bar\xi_i}{q_3}
\end{equation}

The first two terms of the last derivative are easy too
\begin{equation}
\begin{aligned}
\frac{\partial q_1}{\partial\alpha}&=\frac1{(2-\alpha)^2},\\
\frac{\partial q_2}{\partial\alpha}&=-\frac{Nx_M\ln f}{f^\alpha F^2}.
\end{aligned}\hspace*{6cm}\null
\end{equation}

Finally, the last term $\partial q_3/\partial\alpha$ is solved by making use of the identity
\begin{equation}
f_1(\alpha)^{f_2(\alpha)}=\text{e}^{f_2(\alpha)\ln f_1(\alpha)}
\end{equation}
with $f_1(\alpha)$ and $f_2(\alpha)$ two generic functions of the variable $\alpha$, so that
\begin{equation}
\frac{\text{d}}{\text{d}\alpha}f_1(\alpha)^{f_2(\alpha)}=f_1(\alpha)^{f_2(\alpha)}\left[f_2^\prime(\alpha)\ln f_1(\alpha)+f_2(\alpha)\frac{f_1^\prime(\alpha)}{f_1(\alpha)}\right].
\end{equation}
Then
\begin{align}
\frac{\partial q_3}{\partial\alpha}=&\frac{t_i^\frac{2-\alpha}{1-\alpha}}{1-\alpha}\left[\frac{\ln t_i}{1-\alpha}-\frac{(2-\alpha)if^{1-\alpha}\ln f}{Nt_i}\right]+\\\nonumber
&-\frac{t_{i-1}^\frac{2-\alpha}{1-\alpha}}{1-\alpha}\left[\frac{\ln t_{i-1}}{1-\alpha}-\frac{(2-\alpha)(i-1)f^{1-\alpha}\ln f}{Nt_{i-1}}\right].
\end{align}

In case the routine computing the derivatives is called with $\alpha=1$ or 2, we use the above formulae setting $\alpha=\alpha+10^{-5}$.

\section{Ancillary mathematical demonstrations}\label{ancForm}
\subsection{Asymptotical formulae for particular values of $f$}\label{forF}
Equation~(\ref{aM1}) is the asymptotic form of equation~(\ref{aTr}) in the limit $x_M\rightarrow\infty$ or, since $x_m$ is finite, $f\rightarrow\infty$. To show this, we have to compute
\begin{equation}
\lim_{f\rightarrow\infty}\frac{N\ln f}{1-f^{\alpha_\text{T}-1}}
\end{equation}
which is of the form $+\infty/-\infty$. Through de l’H\^opital rule the limit becomes
\begin{equation}
\lim_{f\rightarrow\infty}\frac{N/f}{-(\alpha_\text{T}-1)f^{\alpha_\text{T}-2}}=\lim_{f\rightarrow\infty}\frac{N}{-(\alpha_\text{T}-1)f^{\alpha_\text{T}-1}}=0\label{limHop}
\end{equation}
because $\alpha_\text{T}>1$.

Thus, in the limit $f\rightarrow\infty$ Equation~(\ref{aTr}) becomes
\begin{equation}
\frac{N}{\alpha_\text{T}-1}=\sum\limits_{i=1}^N\ln\frac{x_i}{x_m}
\end{equation}
that gives
\begin{equation}
\alpha_\text{T}=1+\frac{N}{\sum\limits_{i=1}^N\ln\frac{x_i}{x_m}}=\alpha_\text{S}
\end{equation}
QED.

de l’H\^opital rule is used again to find the value of the second term between square brackets in equation~(\ref{alphaSalphaT}) when $f\rightarrow1$. In this case, in fact, the term is indeterminate in the form $0/0$. As done for equation~\ref{limHop}, we write
\begin{equation}
\lim_{f\rightarrow1}\frac{\ln f}{1-f^{\alpha_\text{T}-1}}=-\frac1{\alpha_\text{T}-1}
\end{equation}
so that the r.h.s. of equation~(\ref{alphaSalphaT}) becomes $1+1/0$ and $\alpha_\text{S}\rightarrow\infty$.

\subsection{The consequence of using a wrong estimate of $x_m$ on $\alpha_\text{S}$}\label{wXM}
If equation~(\ref{aM1}) is computed with a parameter $z\ne x_m$, the exponent $\bar\alpha$ we derive is
\begin{equation}
\bar\alpha=1+\frac{N}{\sum\limits_{i=1}^N\ln\frac{x_i}z}
\end{equation}

The sum of logarithms can be expressed in terms of $x_m$
\begin{equation}
\sum\limits_{i=1}^N\ln\frac{x_i}z=\sum\limits_{i=1}^N\ln\left(\frac{x_i}{x_m}\frac{x_m}z\right)=N\ln\frac{x_m}z+\sum\limits_{i=1}^N\ln\frac{x_i}{x_m}
\end{equation}

From equation~(\ref{aM1}) we have
\begin{equation}
\sum\limits_{i=1}^N\ln\frac{x_i}{x_m}=\frac{N}{\alpha_\text{S}-1}
\end{equation}
so that
\begin{equation}
\bar\alpha=1+\frac{N}{N\ln\frac{x_m}z+\frac{N}{\alpha_\text{S}-1}}
\end{equation}
from which equation~(\ref{zm}) follows.

\section{How to generate simulated data following a power-law distribution}\label{genera}
To obtain a variable $x$ distributed according to equation~(\ref{CM1}) in the interval $x_m\le x<x_M$, we follow \citet{Clauset} who derived $p(x)$ for the case $x_m\le x<\infty$, in terms of $p(r)$ where $r$ is distributed uniformly in the interval $0\le r<1$.

Starting from the cumulative functions of both distributions we can write
\begin{equation}
\frac{\alpha-1}{x_m^{1-\alpha}-x_M^{1-\alpha}}\int_{x_m}^xq^{-\alpha}\text{d}q=\frac{x_m^{1-\alpha}-x^{1-\alpha}}{x_m^{1-\alpha}-x_M^{1-\alpha}}=\int_0^r\text{d}t=r
\end{equation}
from which
\begin{equation}
x=x_m\left[1-r(1-f^{1-\alpha})\right]^\frac1{1-\alpha}.\label{xTFr}
\end{equation}

The formula derived by \citet{Clauset} is the limiting case of equation~(\ref{xTFr}) when $f\rightarrow\infty$
\begin{equation}
x=x_m(1-r)^\frac1{1-\alpha}.\label{xFr}
\end{equation}


\bsp	
\label{lastpage}
\end{document}